\documentclass[prx,aps,superscriptaddress, twocolumn,floatfix,longbibliography]{revtex4-1}
\usepackage[utf8]{inputenc}

\usepackage{natbib}
\usepackage{graphicx}
\usepackage{color}
\usepackage[tbtags]{amsmath}
\usepackage[colorlinks, linkcolor=blue]{hyperref}
\usepackage{amssymb}
\usepackage{gensymb}
\usepackage{float}
\usepackage{amsmath}
\usepackage{tabularx,graphicx}
\usepackage{epstopdf}
\usepackage{latexsym}
\usepackage{color, colortbl}
\usepackage{psfrag}
\usepackage{bbm}
\usepackage{bm}
\usepackage{titlesec}
\usepackage{dsfont}
\usepackage{feynmp}
\usepackage{slashed}
\usepackage{multirow}
\usepackage{subfigure}
\usepackage{appendix}
\usepackage{bbold}

\newcommand{\be}{\begin{equation}}
\newcommand{\ee}{\end{equation}}
\newcommand{\beq}{\begin{eqnarray}}
\newcommand{\eeq}{\end{eqnarray}}
\newcommand{\ba}{\[\begin{aligned}}
\newcommand{\ea}{\end{aligned}\]}

\renewcommand{\Im}{{\rm Im\,}}
\renewcommand{\vec}[1]{{\bf #1}}
\renewcommand{\hat}[1]{{\bf {\widehat #1}}}
\renewcommand{\phi}{\varphi}
\renewcommand{\epsilon}{\varepsilon}
\renewcommand{\dag}{\dagger}

\renewcommand{\vec}[1]{\boldsymbol{#1}}

\def \a{{\alpha}}

\def \g{{\gamma}}
\def \D{{\Delta}}
\def \d{{\delta}}
\def \w{{\omega}}

\def \G{{\Gamma}}

\def \ba{\begin{align*}}
\def \ea{\end{align*}}

\newcounter{indice}

\def \bs{\boldsymbol}
\def \mc{\mathcal}

\begin{document}

\title{Robust Gapless Superconductivity in 4Hb-TaS$_2$}
\author{David Dentelski}
\affiliation{Department of Physics, Bar-Ilan University, 52900, Ramat Gan, Israel}
\affiliation{Center for Quantum Entanglement Science and Technology, Bar-Ilan University, 52900, Ramat Gan Israel}
\author{Ezra Day-Roberts}
\affiliation{School of Physics and Astronomy, University of Minnesota, Minneapolis, MN 55455, USA}
\author{Turan Birol}
\affiliation{Department of Chemical Engineering and Materials Science,
University of Minnesota, Minneapolis, 55455 MN}
\author{Rafael M. Fernandes}
\affiliation{School of Physics and Astronomy, University of Minnesota, Minneapolis, MN 55455, USA}
\author{Jonathan Ruhman$^{1,\,2}$}

\date{\today}

\begin{abstract}
The superconducting TMD 4Hb-TaS$_2$  consists of alternating layers of H and T structures, which in their bulk form are metallic and Mott-insulating, respectively. 
Recently, this compound has been proposed as a candidate chiral superconductor, due to an observed enhancement of the muon spin relaxation at $T_c$. 4Hb-TaS$_2$ also exhibits a puzzling $T$-linear specific heat at low temperatures, which is unlikely to be caused by disorder. Elucidating the origin of this behavior is an essential step in discerning the true nature of the superconducting ground state. Here, we propose a simple model that attributes the $T$-linear specific heat to the emergence of a robust multi-band gapless superconducting state. We show that an extended regime of gapless superconductivity naturally appears when the pair-breaking scattering rate on distinct Fermi-surface pockets differs significantly, and the pairing interaction is predominantly intra-pocket. Using a tight-binding model derived from first-principle calculations, we show that the pair-breaking scattering rate promoted by slow magnetic fluctuations on the T layers, which arise from proximity to a Mott transition, can be significantly different in the various H-layer dominated Fermi pockets depending on their hybridization with T-layer states. Thus, our results suggest that the ground state of 4Hb-TaS$_2$ consists of Fermi pockets displaying gapless superconductivity, which are shunted by superconducting Fermi pockets that are nearly decoupled from the T-layers. 
\end{abstract}

\maketitle

\section{Introduction}
The observation of superconductivity in transition metal dichalcogenides (TMD), including in monolayer form, has recently spurred significant interest from the community~\cite{sipos2008mott,efetov2016specular,navarro2016enhanced,xi2016ising,ugeda2016characterization,hsu2017topological,yang2018enhanced,lu2015evidence,fatemi2018electrically,dvir2018spectroscopy}. Among them, bulk 4Hb-TaS$_2$ has emerged as a possible unconventional superconductor.
Recent muon-spin rotation ($\mu$SR) measurements performed on 4Hb-TaS$_2$ reveal a slight enhancement of the  relaxation rate, which onsets at the superconducting transition temperature $T_c$ \cite{ribak2020chiral}. The authors (including one of us) have interpreted this signal as evidence of chiral superconductivity. 
However, the correspondence between enhanced muon-relaxation in a superconductor and time-reversal symmetry-breaking is called for reexamination by recent nuclear magnetic resonance (NMR) measurements~\cite{pustogow2019constraints, petsch2020reduction} performed on another candidate chiral superconductors, Sr$_2$RuO$_4$.
Indeed, one may consider alternative mechanisms that tie such a weak enhancement of the muon relaxation to the onset of superconductivity -- e.g. if local magnetic moments are present, which are screened in the metallic state but not below $T_c$. For this reason, it is essential to understand the correct microscopic model and corresponding ground state of superconducting 4Hb-TaS$_2$. 

\begin{figure}[htp]
	\centering
	\includegraphics[width=0.2\textwidth]{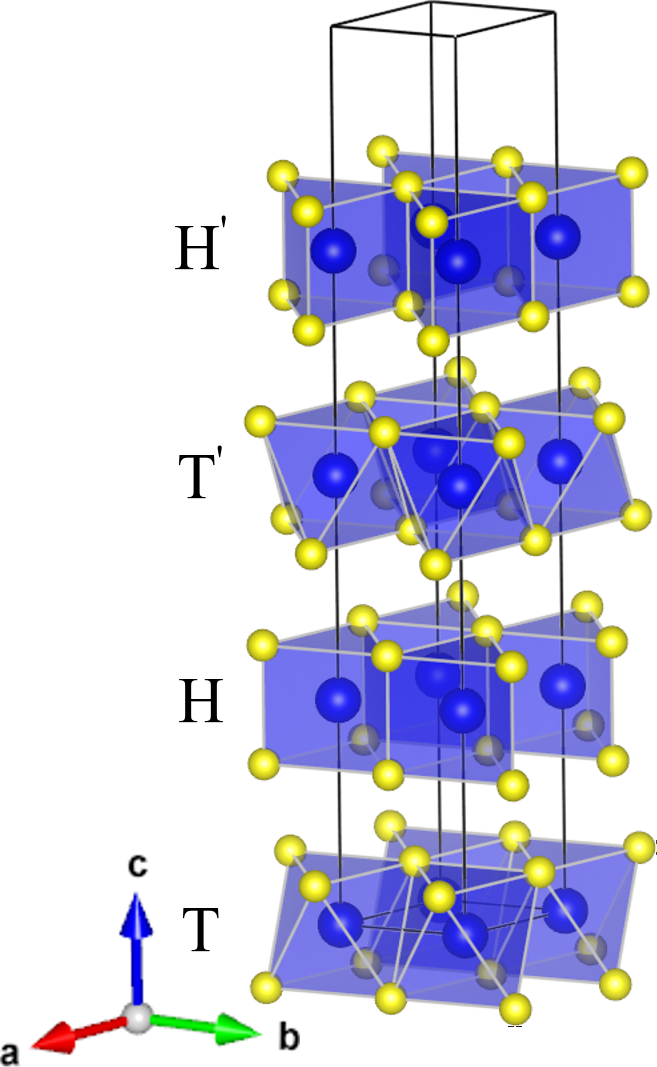}
	\caption{\label{fig:crystal}
		Crystal structure of 4Hb-TaS$_2$. The unit cell consists of four layers of Ta atoms with alternating trigonal (H) and octahedral (T) coordination. Each layer is a triangular lattice of Ta atoms.}
\end{figure}

4Hb-TaS$_2$ is a TMD comprised of alternating single layers with 1T and 1H structures (see Fig. \ref{fig:crystal}). At low temperature, bulk 1T-TaS$_2$ is a strongly correlated insulator due to the small bandwidth associated with a nearly flat conduction band. This flat band results from the reconstruction of the band structure by a prior transition into a ``star-of-David'' charge-density wave (CDW) state \cite{sipos2008mott}. Interestingly, the insulating 1T compound fails to order magnetically to the lowest measured temperatures \cite{fazekas1980charge}, which has motivated the proposal that it may be a spin-liquid \cite{law20171t,ribak2017gapless, murayama2018coexisting} or a dimerized band-insulator~\cite{wang2020band}. 
The 2H polymorph also undergoes a CDW transition, but with a different structure. As apposed to the 1T polymorph, it remains a metal, which is characterized by multiple Fermi pockets. At $T_c^{\rm{2H}}=0.7$K it undergoes a superconducting transition. 

ARPES measurements performed on the 4Hb polymorph show that the star-of-David CDW still forms in the 1T layer~\cite{ribak2020chiral}. However, there is no signature of the insulating state to the lowest measurable temperatures~\cite{disalvo73preparation}. Instead, the entire compound becomes superconducting at $T_c^{\rm{4Hb}} = 2.7$K~\cite{ribak2020chiral}, which is elevated when compared to bulk 2H-TaS$_2$, but still smaller than exfoliated single-layer~\cite{navarro2016enhanced,yang2018enhanced} or intercalated samples~\cite{thompson1972effects}.  Thus, the 4Hb compound intrinsically couples highly-itinerant and nearly localized electrons, a situation that is interesting on its own~\cite{coleman2007heavy}.

A puzzling experimental observation, which may shed light on the ground state properties of 4Hb-TaS$_2$, is the linear in $T$ behavior of the specific heat at low temperatures inside the superconducting state. While the specific heat near the transition temperature $T_c^{\rm{4Hb}}$ resembles that of a fully gapped s-wave superconductor, the low-temperature behavior resembles that of a metal, with a Sommerfeld coefficient value that is $15\%$ of the normal state value \cite{ribak2020chiral}. 
We also note that similar anomalies in the specific heat have been observed in other candidate chiral superconductors, such as the heavy-fermion compound UTe$_2$~\cite{ran2019nearly} and La$_7$Ni$_3$~\cite{singh2020time}. 

The simplest possible explanation for the $T$-linear behavior of the specific heat would be the existence of a sizable non-superconducting volume fraction due to inhomogeneities or impurities even below $T_c$. However, this is unlikely to be the sole explanation, given  the sharp superconducting transition and  homogeneous diamagentic response seen in scanning SQUID measurements   ~\cite{PerskyPrivate}. 
A more exotic explanation would be the emergence of Bogoliubov Fermi surfaces, which are known to arise from time-reversal symmetry (TRS) breaking in nodal superconductors~\cite{agterberg2017bogoliubov} or from pure pair-density wave (PDW) states. However, before reaching the linear-in-$T$ behavior, the specific heat decays exponentially rather than in a power-law fashion, which is inconsistent with a nodal gap (see Appendix~\ref{app:nodal}).
Another option is that strong pair-breaking disorder is present, filling the gap with electronic states. Such a gapless state was predicted by Abrikosov and Gor'kov in their seminal work~\cite{abrikosov1960contribution}. However, in the standard Abrikosov-Gor'kov theory for single-band superconductors, this gapless superconducting state appears only in a very narrow range of the pair-breaking scattering rate, just before superconductivity is completely destroyed by disorder. Moreover, this state is also characterized by a smeared transition and by the absence of coherence peaks in the density of states \cite{skalski1964properties, woolf1965effect}.

In this paper, we apply Abrikosov-Gor'kov theory to the case of two-band superconductors with dominant intra-band pairing interaction. We show that gapless superconductivity naturally emerges over a much wider range of pair-breaking scattering rates when the disorder potential is significantly larger on one of the Fermi pockets. Using a two-band  toy model with these characteristics, we are able to reproduce the temperature-dependence of the specific heat data of 4Hb-TaS$_2$, including the $T$-linear behavior, over a wide range of pair-breaking scattering rates. 

To demonstrate the relevance of this simple model to 4Hb-TaS$_2$, we performed {\it ab-initio}  band structure calculations and derived a simplified tight-binding model using the maximally localized Wannier functions. The Fermi surface consists of several pockets centered at the high-symmetry points of the Brillouin zone. While the spectral weights of the pockets centered at $\G$ and $K$ are dominated by the H-layer states, the $M$-centered pockets have large contributions from the T-layer states. Around each of these high-symmetry points, the Fermi pockets come in pairs -- except at the boundaries of the Brillouin zone at $k_z = \pm \pi/c$, where a screw symmetry makes all bands four-fold degenerate. Since the 2H polymorph superconducts on its own, we associate a pair of pockets with dominant H-layer character in 4Hb-TaS$_2$ to the two bands in our toy model.

The key point is that one of the pockets in each of these pairs hybridizes much more strongly with the T-layer states, particularly at the zone center plane ($k_z=0$). Now, the 1T polymorph on its own is a Mott insulator with presumably strong (but possibly frustrated) magnetic fluctuations arising from local moments. Low-energy, slow magnetic fluctuations are expected to have a pair-breaking effect \cite{Millis1988,Fernandes2013}, which shares some similarities with the pair-breaking effect of magnetic impurities. Therefore, the pocket in the pair that is most strongly hybridized with the T-layer states is expected to be subjected to a stronger pair-breaking potential. We demonstrate that this is indeed the case by computing the scattering rate for each Fermi pocket after modeling the slow magnetic fluctuations as local magnetic ``impurity" scatterers in the T-layer. The scattering rates in each pocket of the same pair is notably different near the $k_z=0$ plane, which qualitatively justifies the assumptions of our toy model.  
Additionally, we observe a significant difference in the scattering strength between the $\G$ and $K$ pockets, where the former is larger.  
Overall, our results suggest that the ground state of 4Hb-TaS$_2$ is a combination of gapless Fermi pockets and fully gapped superconducting pockets, which does not necessarily break time-reversal symmetry. 

The rest of this paper is organized as follows. In Sec.~\ref{Toy Model} we present our two-band toy model, where we assume that only one of the bands is coupled strongly to magnetic moments originated in the T-layer. We show that within this model, gapless superconductivity naturally emerges over a wide range of pair-breaking scattering rates.
To relate our model to the microscopic properties of 4Hb-TaS$_{2}$, 
in Sec.~\ref{TBM} we construct a tight binding  model for this compound. We find that our assumptions of the different coupling to T-layer magnetic moments holds at $k_{z}=0$, where we can expect local moments on the T-layer to couple much weaker to the inner Fermi surfaces in comparison to the outer ones. In Sec.~\ref{Scattering-Rate} we use these results to numerically estimate the scattering rate on the inner and outer Fermi surfaces. 
Sec.~\ref{Discussion} summarizes our results and discusses the limitations and implication of our model to understand the superconducting state of 4Hb-TaS$_{2}$.

\section{Robust gapless superconductivity in a two-band toy-model }\label{Toy Model}

In this section we present a simple toy model to explain the emergence of robust gapless superconductivity in 4Hb-TaS$_2$. We show that such a state is stabilized over a wide range of parameters when one of the two Fermi pockets couples strongly to pair-breaking disorder, while the  other one is only weakly coupled. 

\subsection{Self-consistent gap equation}

We consider the simplest possible scenario of two pockets (labeled 1 and 2) with zero center-of-mass momentum pairing, which are described by the Gor'kov Green's function
\begin{align}\label{eq:G0}
\mathcal {G}_{0}(\textbf{k}, i\omega_{n})= \begin{pmatrix} {G}_{1}(\textbf{k}, i\omega_{n})& 0 \\ 0 & {G}_{2}(\textbf{k}, i\omega_{n}) \end{pmatrix}\,,
\end{align}
written in the Nambu space $\Psi_{\bs k}= (c^\dag_{\bs k\uparrow  1},c_{-\bs k\downarrow  1},c^\dag_{\bs k\uparrow 2},c_{-\bs k\downarrow 2})$, where
\begin{align}\label{eq:Gj}
\begin{split}
{G}_{j}(\textbf{k}, i\omega_{n}) =- \dfrac{i{\omega}_{n}\tau^{0}+\xi_{\textbf{k},j}\tau^{3}+{\Delta}_{j}\tau^{1}}{{\omega}_{n}^{2}+\xi_{\textbf{k},j}^{2}+{\Delta}_{j}^{2}}.
\end{split} 
\end{align}

Here, $\tau^{\a}$ are Pauli matrices in the particle-hole basis, $\xi_{\textbf{k},j}$ is the dispersion relation of band $j=1,2$, $\omega_{n} = 2\pi T (n+1/2)$ are the fermionic Matsubara frequencies, and $T$ is the temperature.  
For simplicity, we assume s-wave pairing, although our main result should hold for any {\it nodeless} gap function (including a chiral state in two-dimensions).  

To determine the magnitude of the superconducting order parameter $\Delta_j$ within Abrikosov-Gor'kov (AG) theory, we first incorporate the effect of disorder by computing the self-energy correction to the Gor'kov Green's function 
\begin{align} \label{eq:SE}
{\Sigma}(\textbf{k},i\omega_n) = n_{imp}\int \dfrac{d^{3}p}{(2\pi)^3} \hat{V}_{\textbf{p}-\textbf{k}}\,\mathcal{G}(\textbf{p},i\omega_{n})\,\hat{V}_{\textbf{k}-\textbf{p}},
\end{align}
where ${\mathcal G}^{-1}={\mathcal G}^{-1}_0-{ \Sigma}$ is the dressed Green's function. 
This correction is obtained  following the standard AG theory, where we average over uncorrelated configurations of point-like defects with concentration  $n_{imp}$. 
We assume that the source of pair-breaking is magnetic disorder in the T-layers. This leads to the Nambu-space disorder-potential matrix
\be
\hat{V}  =  \begin{pmatrix} V_1 & 0 &V_{12} &0 \\ 0 & V_1 & 0 &V_{12} \\ V_{12} &0 &V_2 &0 \\0&V_{12} & 0 & V_2  \end{pmatrix}
\ee
Here $V_{1}$, $V_2$ and $V_{12}$ are the intra- and inter-band disorder potential strengths, respectively. In what follows we assume $V_1, V_{12} \ll V_2$, which implies that band 2 is coupled much stronger to the magnetic impurities than band 1. We will show that in this limit, gapless superconductivity is stabilized over a wide range of parameters. For simplicity, however, let us consider the extreme limit where $V_1 = V_{12} = 0$ and $V_2 \ne 0$.

In this case, the block of the Green's function corresponding to band 1 retains its form in Eq.~\eqref{eq:G0}, while the block of band 2 needs to be calculated self-consistently using Eq.~\eqref{eq:SE}. 
The solution is obtained by substituting, in the Green's function expression in Eq. (\ref{eq:Gj}), $\Delta_2 \to \tilde \D_2 = \Delta_{2}-{\Gamma}{\tilde{\Delta}_{2}}/{\left( 2\sqrt{\tilde{\omega}^{2}_{n}+\tilde{\Delta}_{2}^{2}}\right)}$ and $\w_n \to \tilde \w_n=\omega_{n}+{\Gamma}{ \tilde{\omega}_{n}}/{\left(2\sqrt{\tilde{\omega}^{2}_{n}+\tilde{\Delta}_{2}^{2}}\right)}$. Here, $\Gamma=2\pi V^{2} n_{imp}  \nu_2$ is the pair-breaking scattering rate and $\nu_2$ is the density of states of band 2. 
These equations can be written in a concise manner by defining $x_n ={\tilde{\Delta}_{2}}/{\tilde{\omega}_{n}}$ and dividing them by one another, such that
\begin{align}\label{eq:SCx}
x_n  =  \dfrac{\Delta_{2}-{\Gamma|x_n|/ \left( 2 \sqrt{1+x_n^{2}}\right)}}{\omega_{n}+{\Gamma  {\rm{sign}}(\omega_{n})/ \left( 2\sqrt{1+x_n^{2}}\right)}}.
\end{align}

We are now in position to determine the magnitude of  $\Delta_j$ from the gap equation using the dressed Green's function
\begin{align} \label{eq:SCdelta}
\begin{pmatrix} \Delta_{1}\\ \Delta_{2} \end{pmatrix} =T \pi\sum_{n} \begin{pmatrix} \lambda_{1} & u \\ u &  \lambda_{2}  \end{pmatrix} \begin{pmatrix}  {{\Delta}_{1}\over \sqrt{{\omega}^{2}_{n}+{\Delta}_{1}^{2}}}\\ {\tilde{\Delta}_{2}\over \sqrt{\tilde{\omega}^{2}_{n}+\tilde{\Delta}_{2}^{2}}} \end{pmatrix},
\end{align}
where $\lambda_{j}$ and $u$ are dimensionless intraband and interband pairing couplings, respectively.

Eqs.~(\ref{eq:SCx}), (\ref{eq:SCdelta}) can be readily solved numerically by iteration. However, it is also helpful to obtain asymptotic solutions in the limits $x_n \gg 1$, corresponding to weak disorder and small frequency, and $x_n \ll 1$, corresponding to strong disorder and large frequency

In the regime $x_n \gg 1$ we find the asymptotic solution:
\begin{align} \label{eq:BigXlim}
x_n \sim \dfrac{\Delta_{2}-{\Gamma}}{\omega_{n}}\;,\;\w_n \to 0\,.
\end{align}
This equation reflects the well known result in AG theory that the critical scattering rate above which gapless superconductivity emerges is given by $\G_{*} = \Delta_2$, which is slightly smaller than the critical value $\G_c$ for which superconductivity is destroyed (i.e. when $\Delta_2$ goes to zero). In what follows, we will show explicitly via a computation of the density of states (DOS) that $\G_{*} = \Delta_2$ indeed implies a finite DOS at the Fermi level and, therefore, gapless superconductivity.

On the other hand, for $x_n \ll 1$, the asymptotic solution is given by:
\begin{align}\label{eq:SmallXlim}
x_n \sim \dfrac{\Delta_{2} \ {\rm sign}(\omega_{n})}{|\omega_{n}|+\Gamma}\;,\; \D_2 / \G \ll 1\,.
\end{align}
Note that this solution becomes asymptotically exact in the limit $\D_2 \ll \G$, which defines the strong disorder limit, and in the high frequency limit $\w \to \infty$.

\begin{figure}[tbh]
	\centering
	\begin{subfigure}
		\centering
		\includegraphics[width=1\columnwidth]{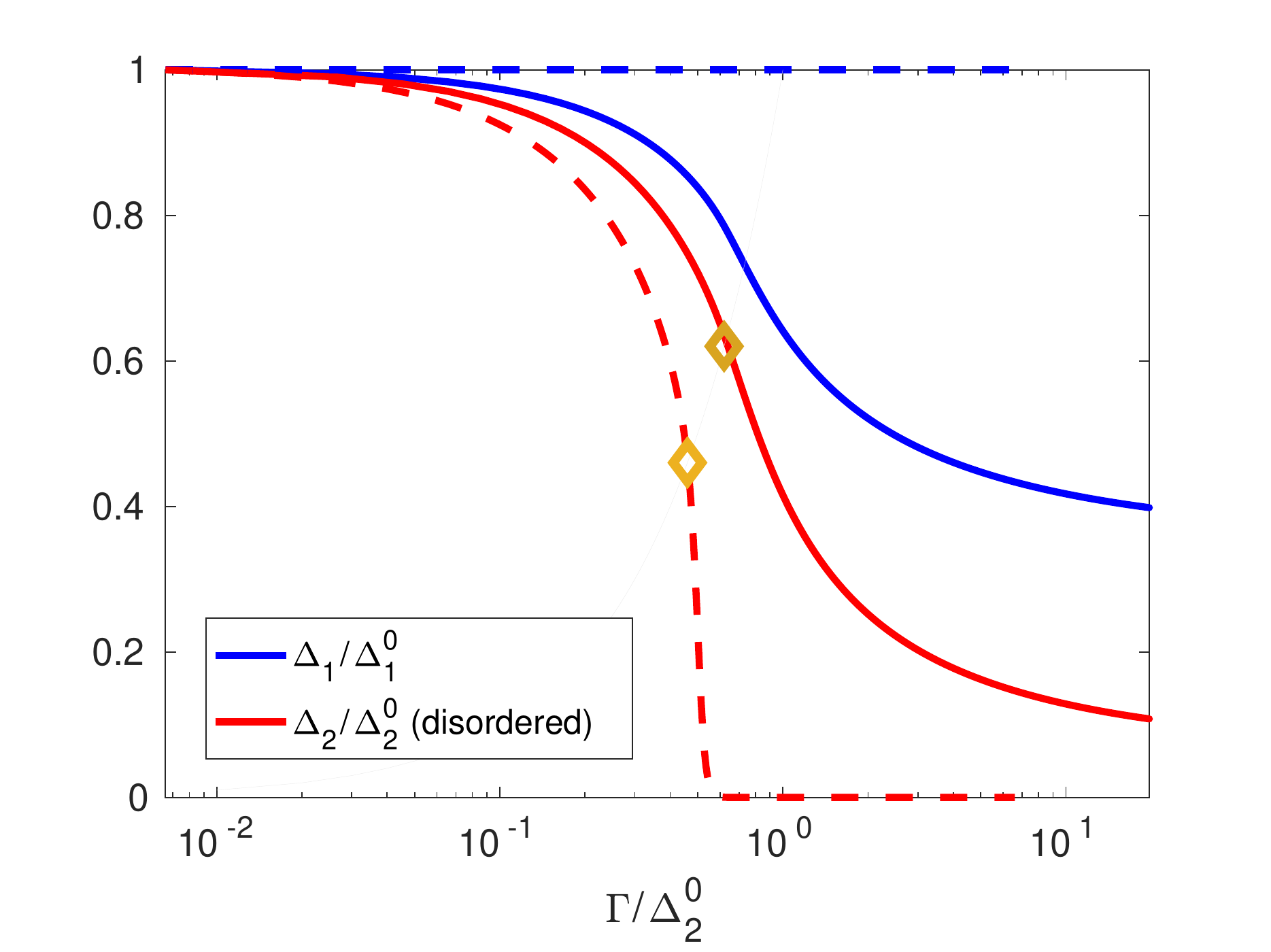}
		\label{fig:fpr}
		(a) 
	\end{subfigure}
	\begin{subfigure}
		\centering
		\includegraphics[width=1\columnwidth]{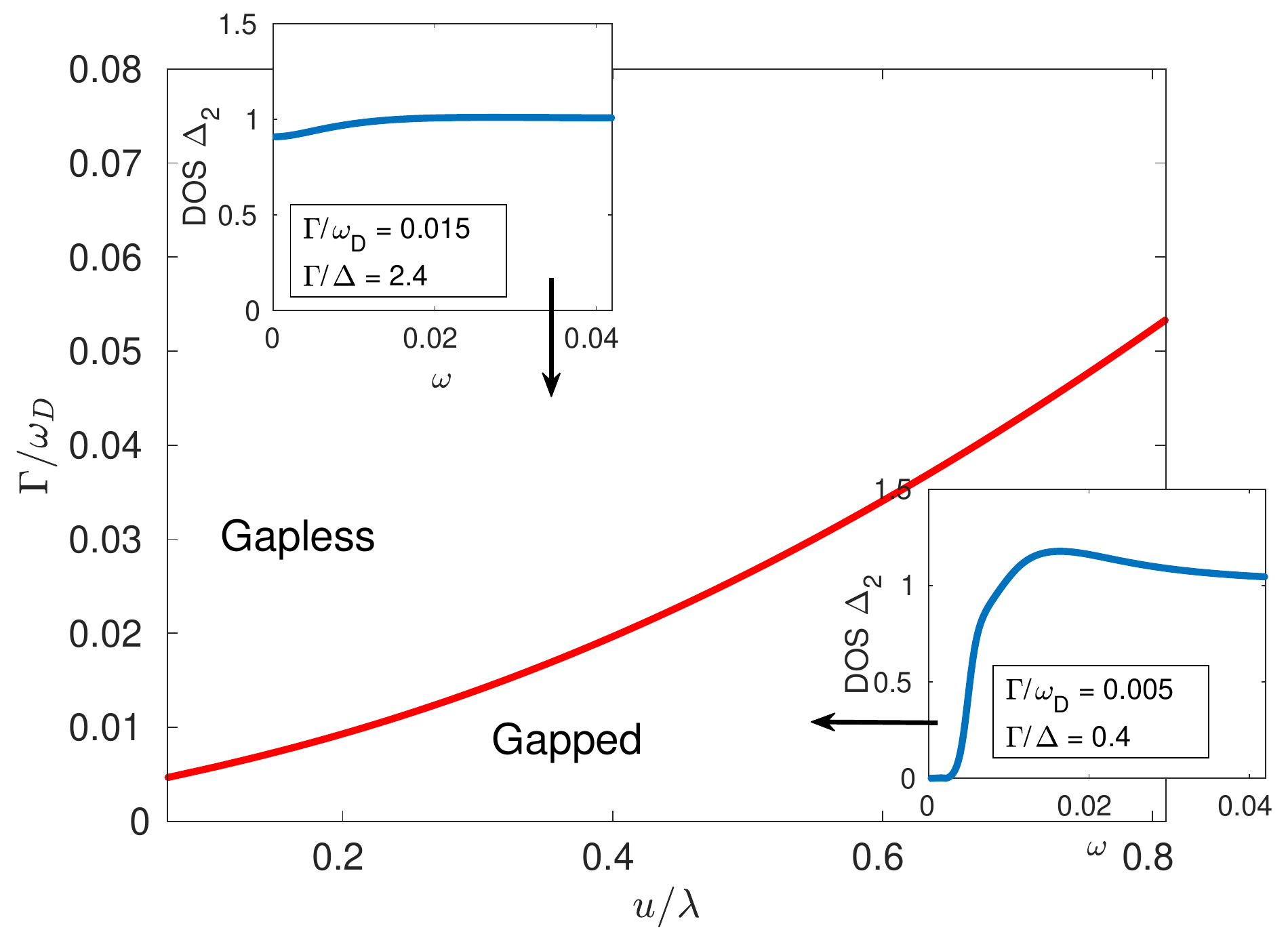}
		\\
		(b)
	\end{subfigure}
	\caption{(a) The superconducting order parameters $\Delta_{1}$ and $\Delta_{2}$, as a function of the pair-breaking rate $\Gamma$ obtained from numerical solutions of Eqs.~(\ref{eq:SCx}), (\ref{eq:SCdelta}). The dashed lines represents the case of zero inter-pocket pairing $u = 0 $ and $\lambda_{1}=\lambda_{2} = -0.2$. In this case, $\Delta_{1}$ is unaffected by disorder while $\Delta_{2}$ goes rapidly to zero. As a result, the region of gapless superconductivity, starting at the yellow diamond, is extremely narrow (the gray line is $\Delta_{2}=\Gamma$ on a logarithmic scale). In contrast, the case of finite inter-pocket coupling $u = -0.035$ and $\lambda_{1} = \lambda_{2} = -0.17$ is represented by the solid lines. In this case, $\Delta_{2}$ decreases slowly with $\Gamma$, resulting in a much wider region of gapless superconductivity.
		(b) The $\Gamma-u$ ``phase diagram" that determines the conditions for gapless and gapped superconductivity in band 2. The inset show the corresponding  density of states (DOS) of band 2 for two representative values of disorder, illustrating the gapped (bottom right) and gapless (top left) superconducting phases.}
	\label{fig:GaplessDOS}
\end{figure}

In both regimes, the value of $\D_2$ is obtained from the self-consistent solution of the gap Eq.~\eqref{eq:SCdelta}. In Fig.~\ref{fig:GaplessDOS}(a) we plot the solution for the two gaps as a function of $\G$ for two cases, one with a finite but small inter-pocket pairing interaction $u=-0.035$ and  $\lambda_1 = \lambda_2 = -0.17$ (solid  lines), and one where $u = 0$ and $\lambda_1 = \lambda_2 = -0.2$ (dashed lines).  

Let us first consider the latter case where there is no inter-pocket pairing. In this situation, $\Delta_1$ is not affected by disorder, while $\D_2$ is suppressed to zero very quickly.
In agreement with AG theory, the gap vanishes slightly after reaching the gapless regime $\Delta_ 2 = \G$ as predicted by Eq.~\eqref{eq:BigXlim} (which is marked by a yellow diamond in the figure).

On the other hand, when $u$ is small but finite there is always a non-zero solution for $\D_2$, which decreases monotonically with $\G$. In particular, the region of gapless superconductivity is extended to a much wider range of $\G$ values, starting from the value corresponding to the yellow diamond  in Fig.~\ref{fig:GaplessDOS}(a) and extending all the way up.
In Fig.~\ref{fig:GaplessDOS}(b), we present the corresponding ``phase diagram" for gapped and gapless superconductivity in band 2 in the parameter space of $\Gamma$ and inter-pocket pairing $u$.  

It should be noted that in any realistic system, band 1 will also be coupled to pair breaking disorder, characterized by some rate $\G_1$. It is also possible that inter-pocket pair-breaking $\G_{12}$ emerges in the case of non-pointlike impurities. The working hypothesis of this paper is that both pair-breaking scattering rates are much smaller than $\G$. Denoting the maximum of the two by $\rm{max}(\G_1,\G_{12})=\d \,\G$, such that $\d \ll 1$, robust gapless superconductivity appears in the wide regime  $\D_2<\G<\D_1/\d$. 

Therefore, we have demonstrated that multi-pocket superconductors with disorder that couples much stronger to one pocket than the other exhibit gapless superconductivity over a wide range of disorder strengths. We now turn to discuss the experimental consequences of such a state, focusing on the specific heat.

\subsection{Specific heat}
The thermodynamic properties of the superconducting state is governed by the fermionic density of states (DOS). In the case of zero center-of-mass momentum pairing, the bands are decoupled at the single-particle level and contribute independently to the total density of states.  
The density of states of band 2 is given by 
\begin{align}\label{eq:DOS}
\dfrac{\nu_{2}(\omega)}{\nu_{2}^{0}} = -\dfrac{1}{\pi}\sum_{\textbf{k}} {\rm Im} G^{R}_2(\textbf{k}, \omega) = {\rm Re} \left[ \dfrac{\omega}{\sqrt{\omega^2-\mc{D}^2(\omega)}}\right],
\end{align}
where $G^{R}_2$ is the retarded Green's function of band 2 obtained from analytical continuation, $\nu^{0}_2$ is the density of states at the absence of superconductivity, and $\mc{D}(\omega)$ is the analytic continuation of  
\be\label{eq:D}
\mc{D}(i\omega_n)\equiv x_n \omega_n.
\ee
Similarly, the density of states of band 1 is given by  $\nu_1(\w)/\nu_{1}^{0} = \rm{Re}\left[ \w / \sqrt{\w^2 - \D_1^2}\right]$. Thus, the total density of states is 
$\nu(\w) = \nu_1(\w) + \nu_2(\w)$.

It is clear from Eqs. (\ref{eq:DOS}) and (\ref{eq:D}) that when $x_n \omega_n = 0$, a finite density of states emerges at the Fermi level. Using Eq. (\ref{eq:BigXlim}), we see that this happens when $\Gamma = \Delta$. In the limit of strong disorder, $\Gamma \gg \Delta$, we can use Eq.~\eqref{eq:SmallXlim} for band 2. Assuming that band 1 is fully gapped, we obtain  a finite density of states at $\w \to 0$ given by
\be
\nu(\w) \sim {\nu_2^0 \over \sqrt{1+ {(\D_2 / \G)^2}}}\;,\;\w\to 0
\ee
which leads to a finite Sommerfeld coefficient $\g = \pi^2 k_B \nu(0)/3$. To compute the precise value of this contribution for arbitrary disorder strength, as well as its evolution with temperature, we resort to the numerical computation. 

\begin{figure*}[tbh]
	\centering
	\begin{subfigure}
		\centering
		\includegraphics[width=0.5\columnwidth]{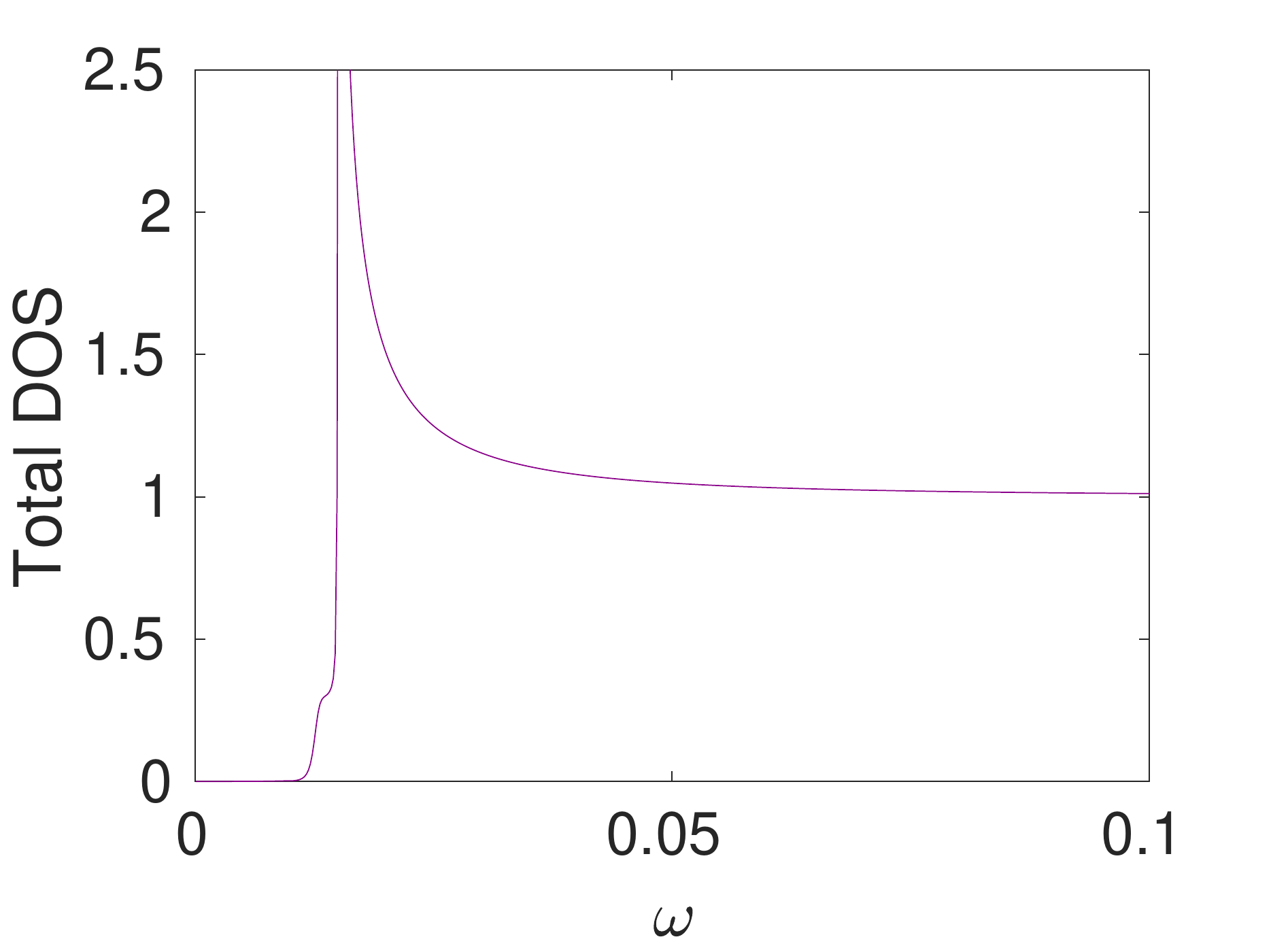}
	\end{subfigure}
	\hfill
	\begin{subfigure}
		\centering
		\includegraphics[width=0.5\columnwidth]{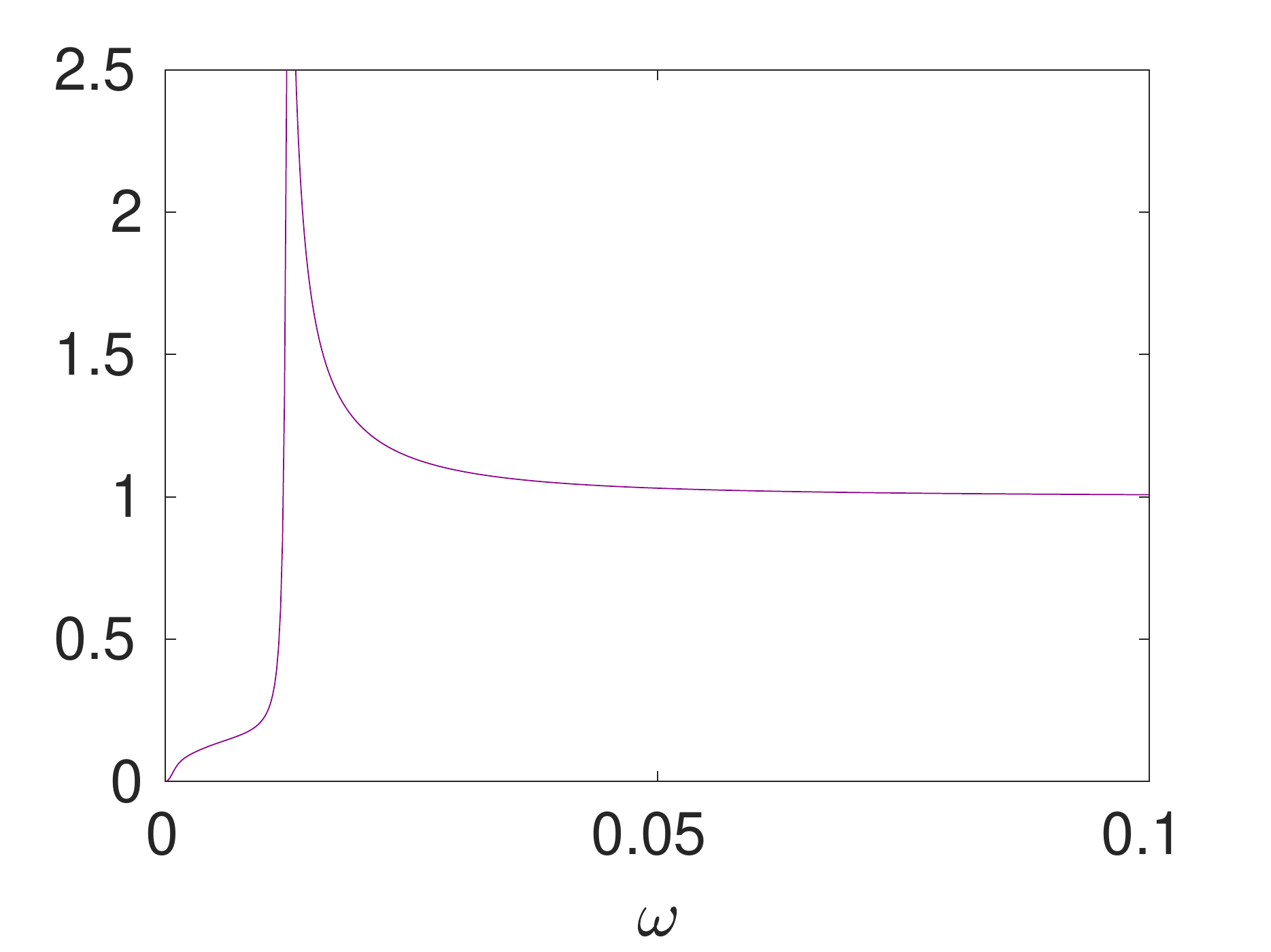}
	\end{subfigure}
	\hfill
	\begin{subfigure}
		\centering
		\includegraphics[width=0.5\columnwidth]{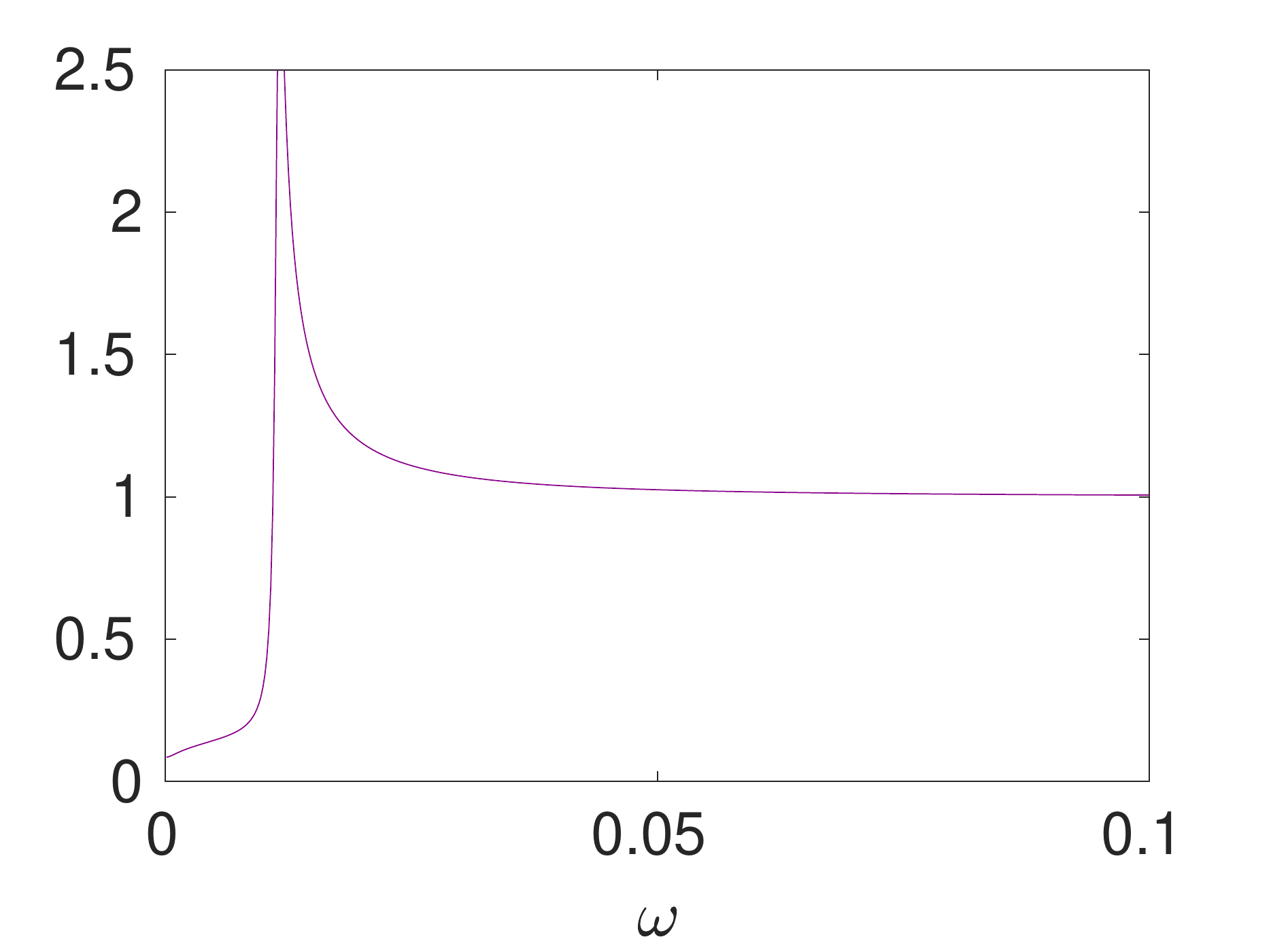}
	\end{subfigure}
	\hfill
	\begin{subfigure}
		\centering
		\includegraphics[width=0.5\columnwidth]{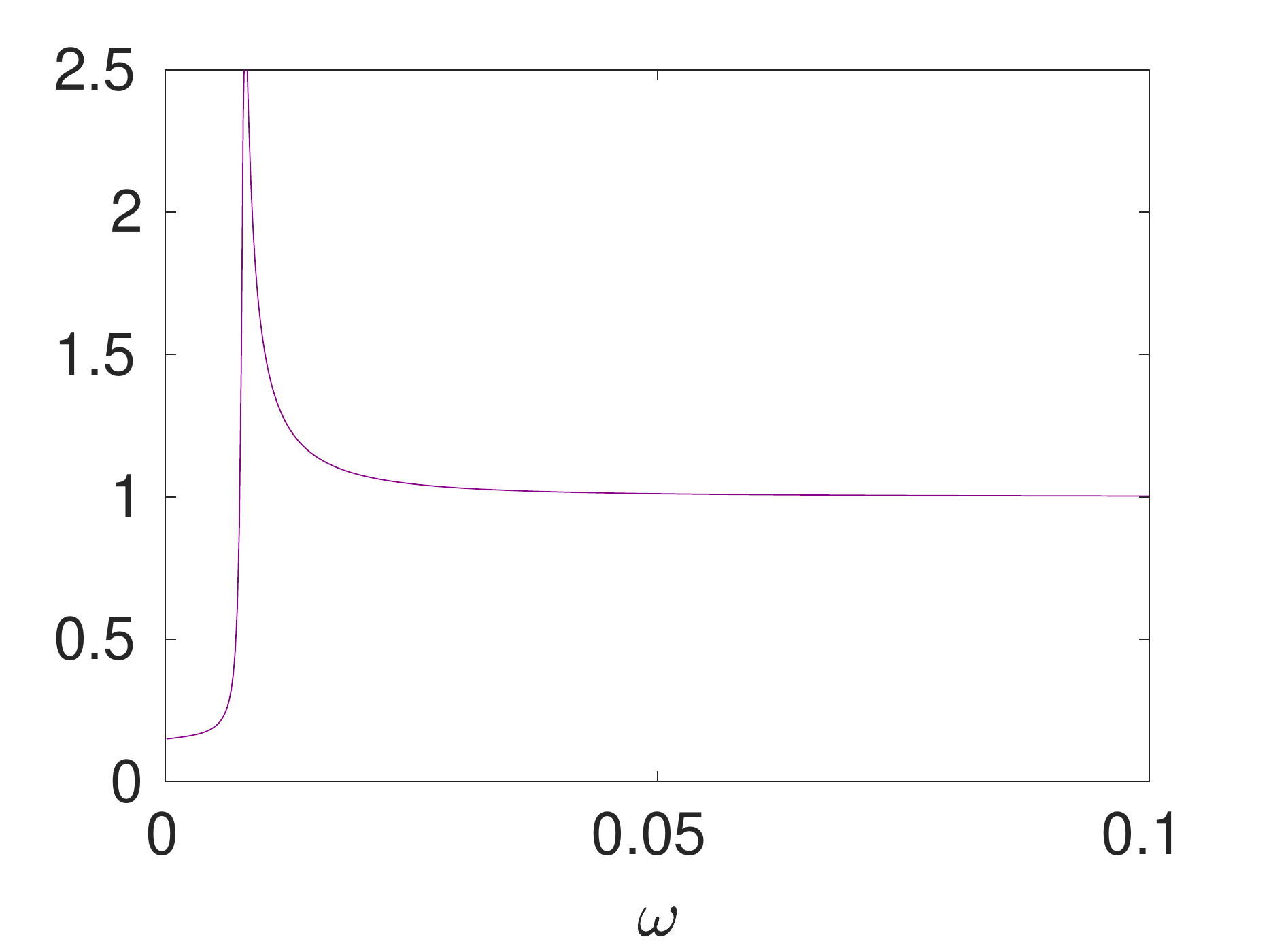}
	\end{subfigure}
	\begin{subfigure}
		\centering
		\includegraphics[width=0.5\columnwidth]{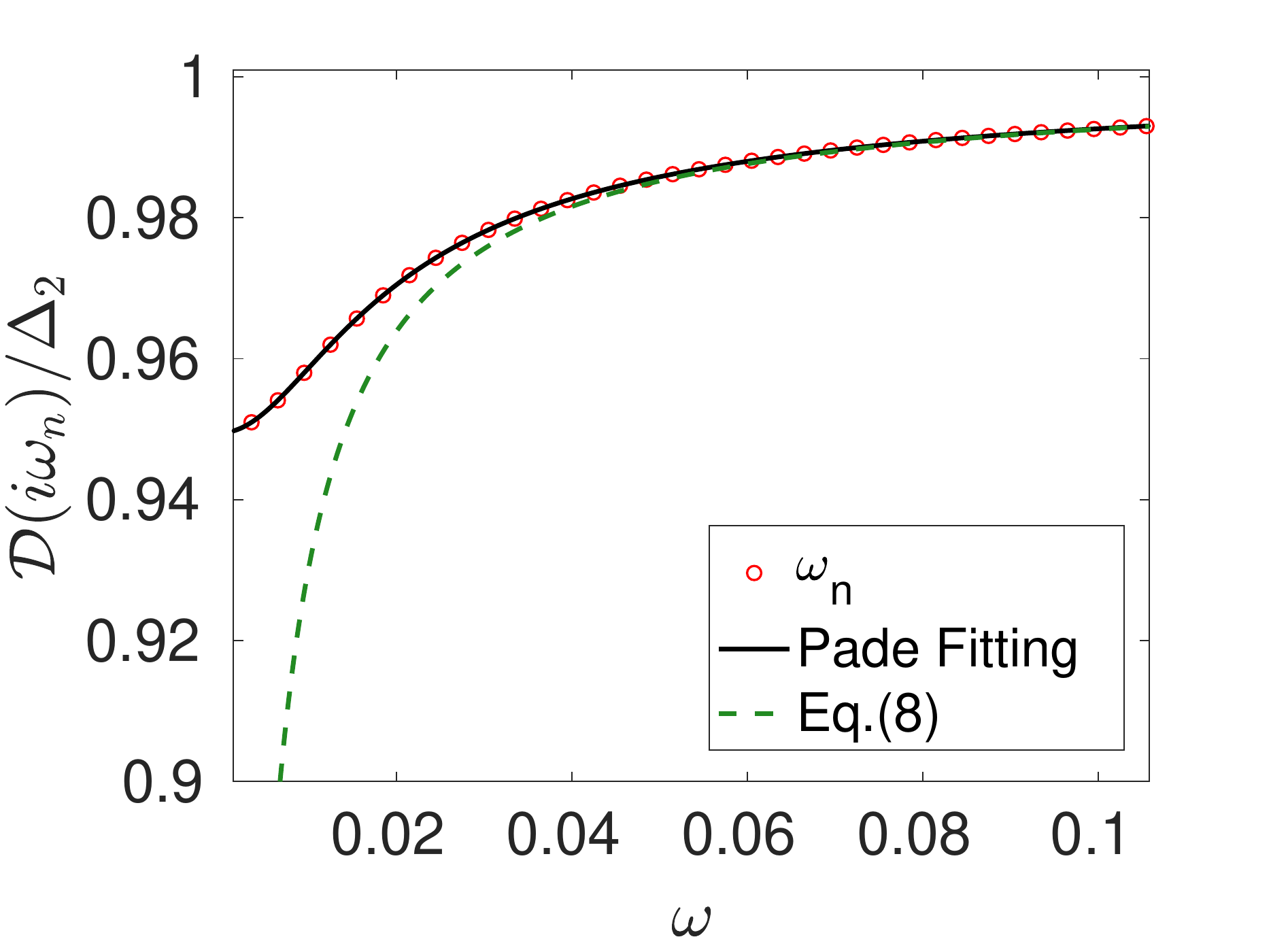}
	\end{subfigure}
	\begin{subfigure}
		\centering
		\includegraphics[width=0.5\columnwidth]{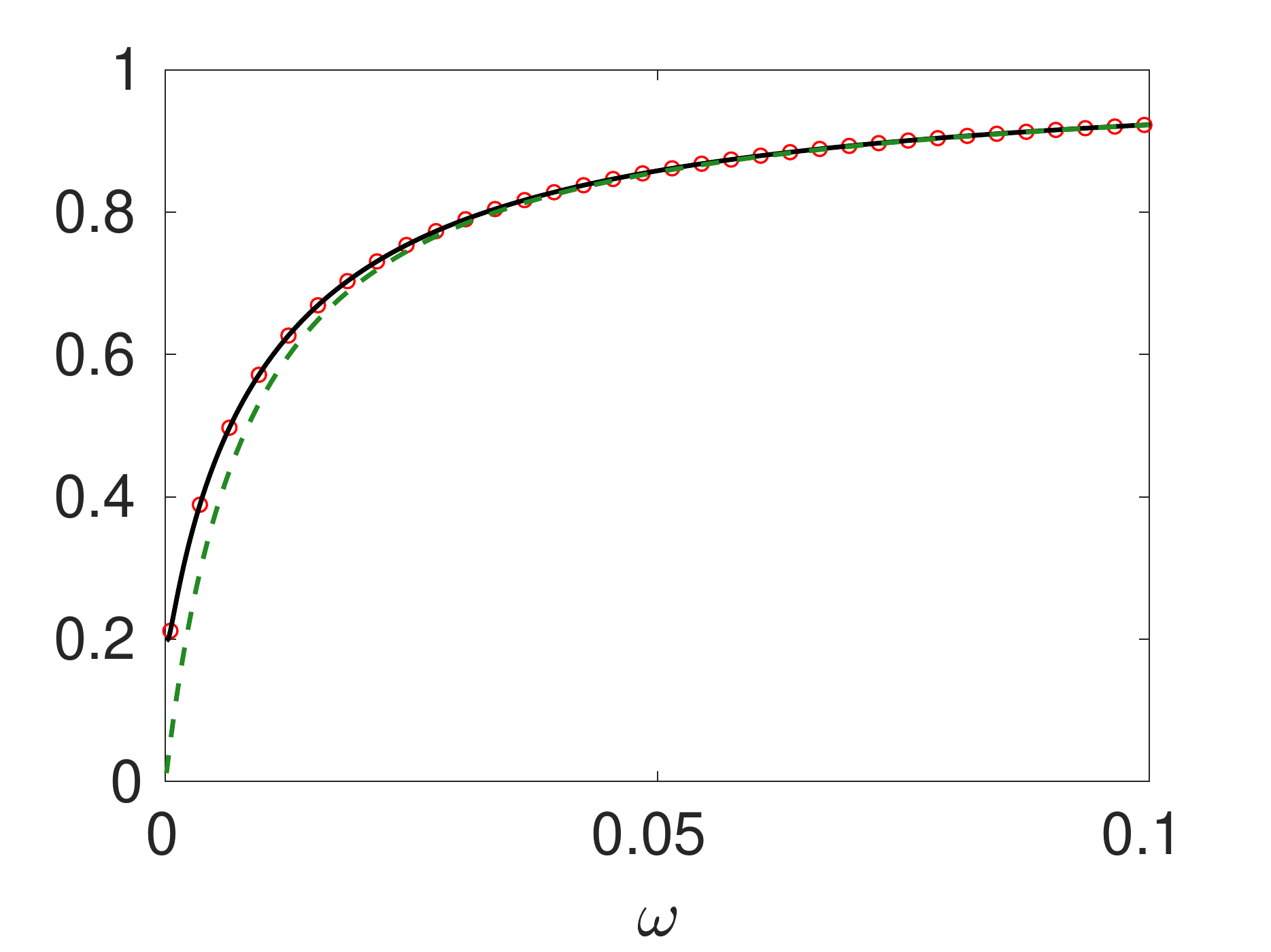}
	\end{subfigure}
	\begin{subfigure}
		\centering
		\includegraphics[width=0.5\columnwidth]{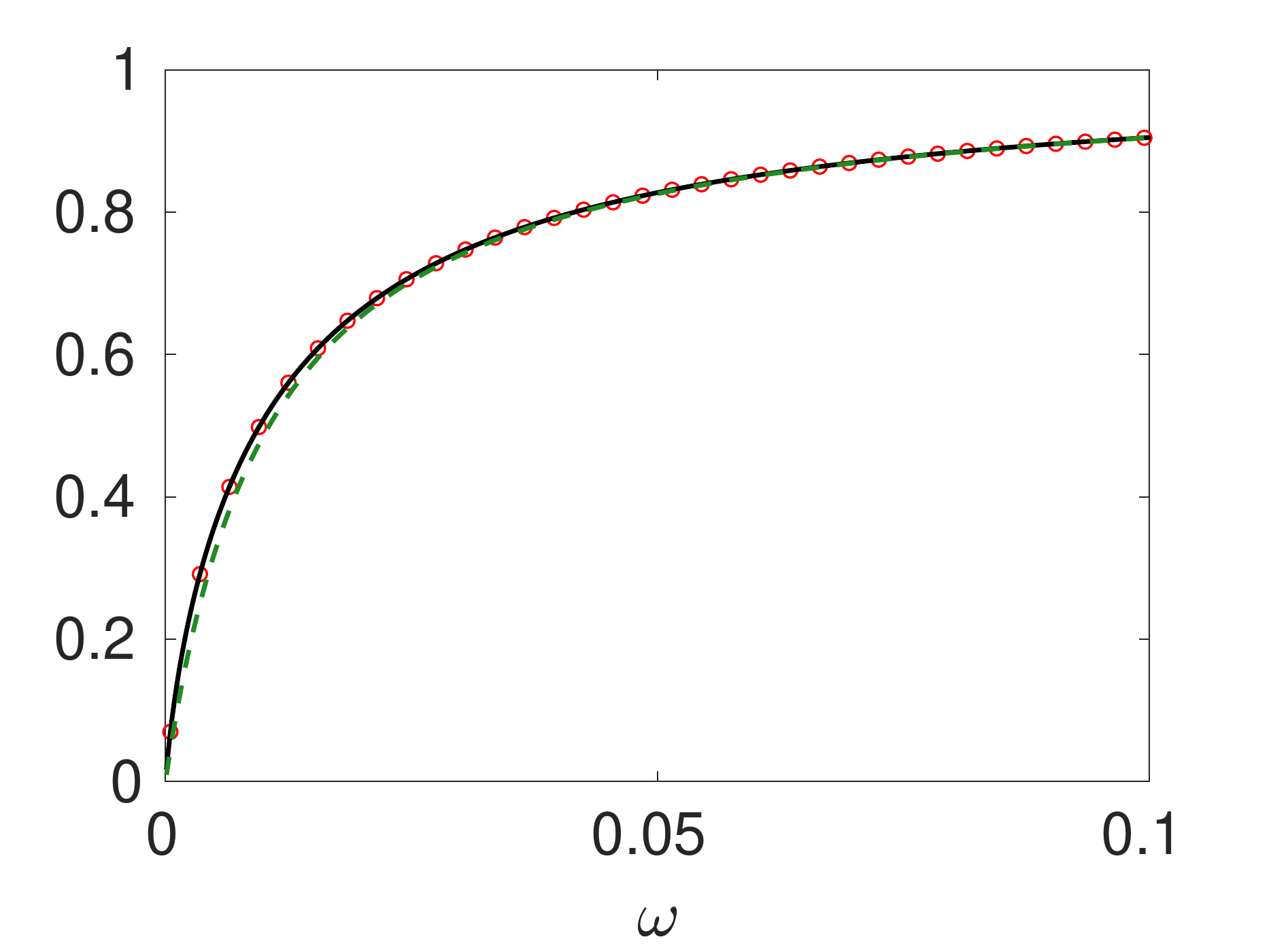}
	\end{subfigure}
	\begin{subfigure}
		\centering
		\includegraphics[width=0.5\columnwidth]{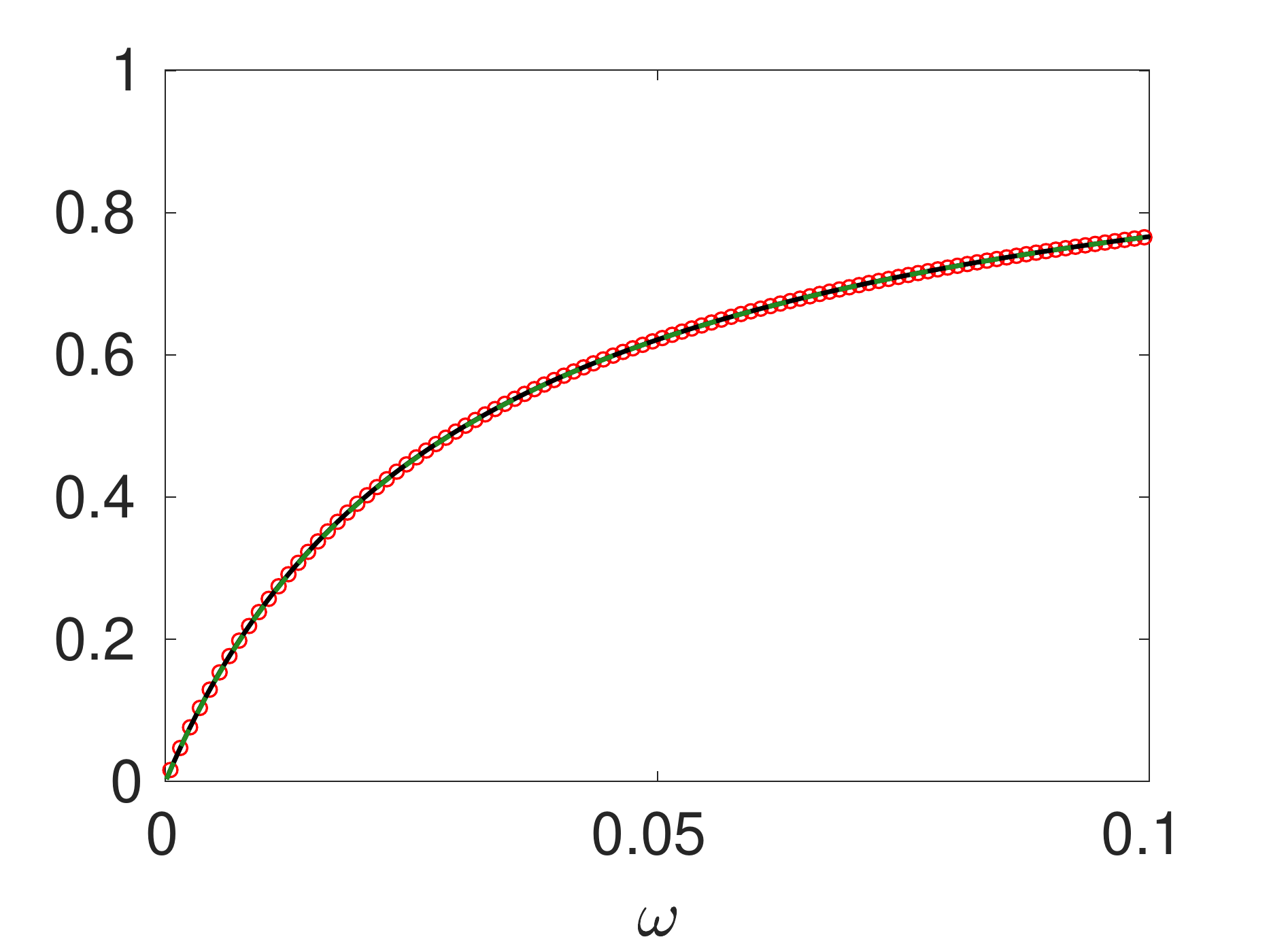}
	\end{subfigure}
	\begin{subfigure}
		\centering
		\includegraphics[width=0.5\columnwidth]{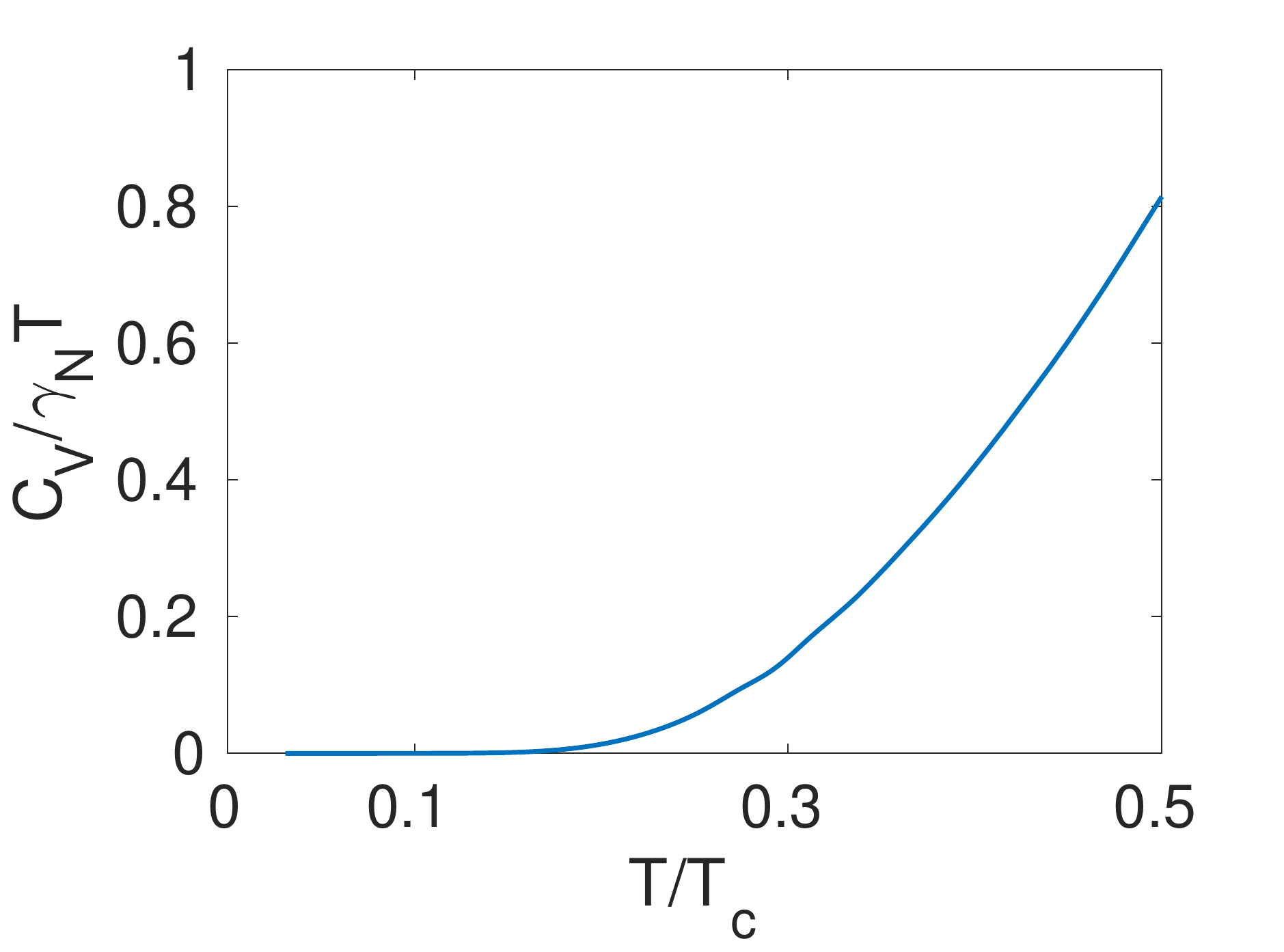}
	\end{subfigure}
	\begin{subfigure}
		\centering
		\includegraphics[width=0.5\columnwidth]{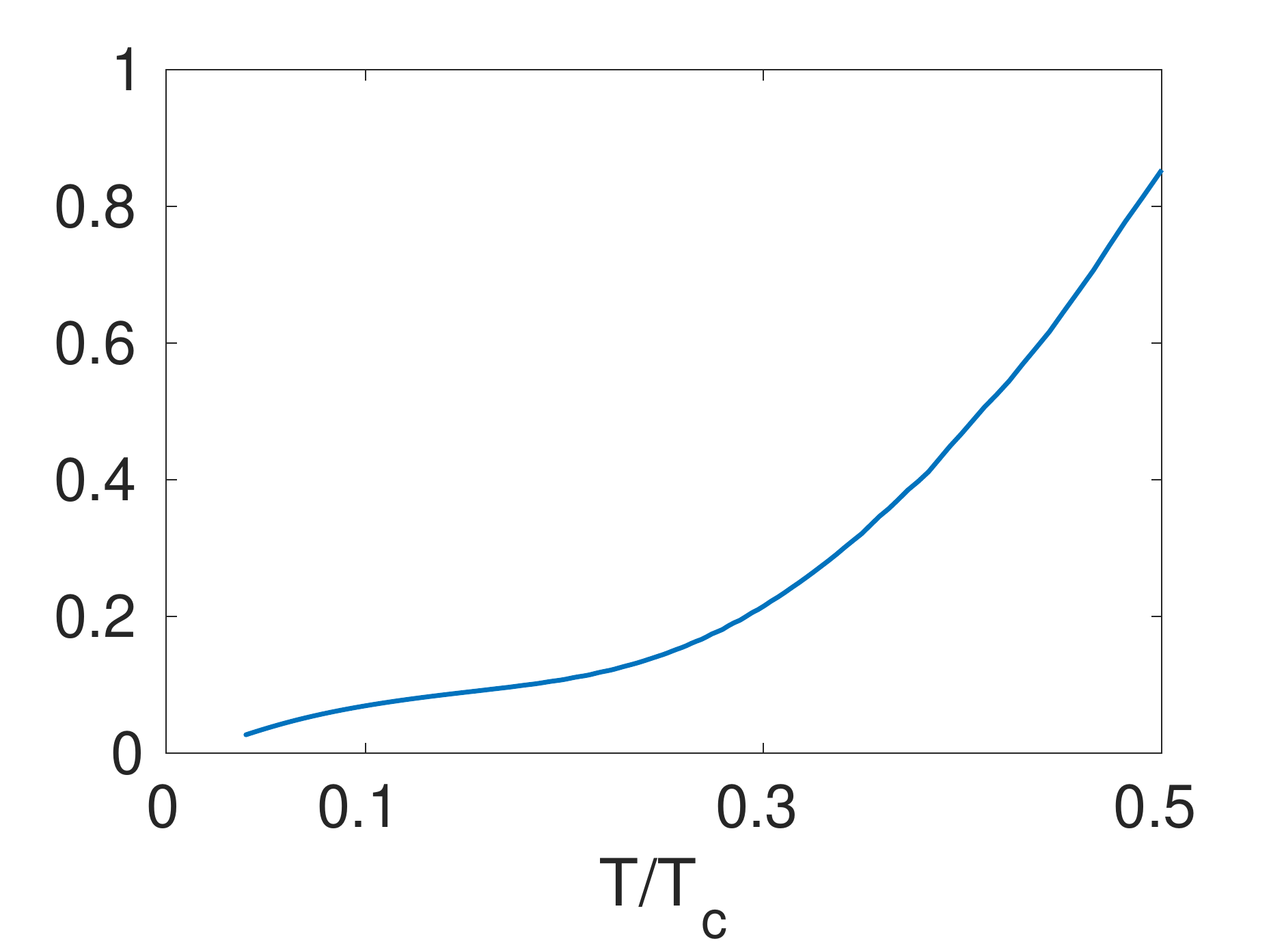}
	\end{subfigure}
	\begin{subfigure}
		\centering
		\includegraphics[width=0.5\columnwidth]{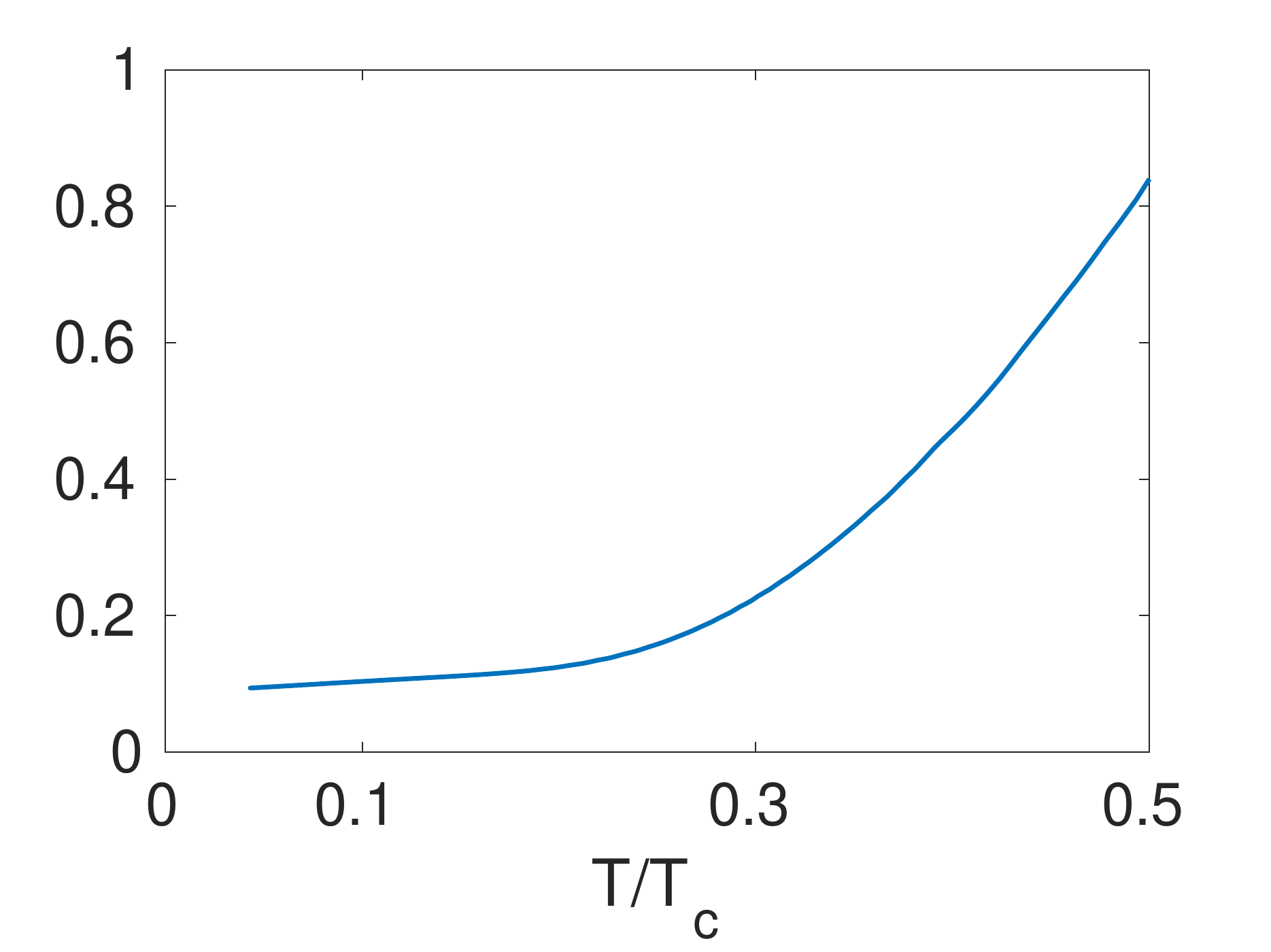}
	\end{subfigure}
	\begin{subfigure}
		\centering
		\includegraphics[width=0.5\columnwidth]{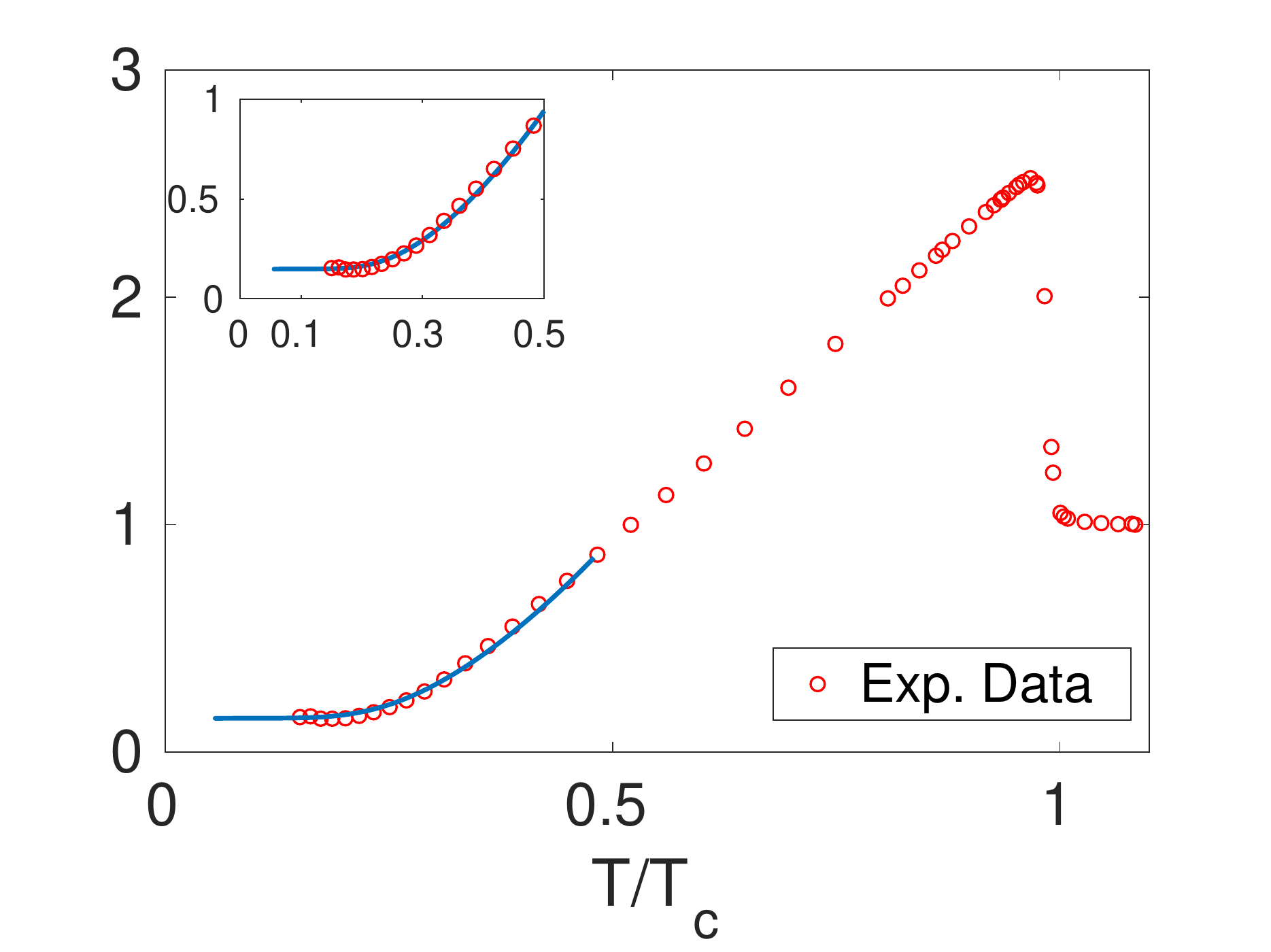}
	\end{subfigure}
	(a) $\Gamma/\Delta_{2} = 0.05$ \ \ \ \ \ \ \ \ \ \ \ \ \ \ \ \  \ \ \ \ \ (b) $\Gamma/\Delta_{2} = 0.8$ \ \ \ \ \ \ \ \ \ \ \ \ \ \ \ \ \ \ \ \ \ (c) $\Gamma/\Delta_{2} = 1.2$ \ \ \ \ \ \ \ \ \ \ \ \ \ \ \ \  \ \ \ \ \  (d) $\Gamma/\Delta_{2} = 8$ \\
	\caption{The first two rows present the normalized tunneling density of states $\nu(\omega)/\nu_{0}$ and the Matsubara-frequency dependent gap function $\mc{D}(i\omega_{n})/\Delta_{2}$. The gap function in the middle row compared with the results from the Pad{\'e} approximation used for the analytical continuation \cite{beach2000reliable} and the asymptotic form at large Matsubara frequency given by Eq.~(\ref{eq:SmallXlim}). Clearly, the asymptotic form captures the entire frequency dependence of the gap in the limit of large $\G$. The last row shows the corresponding heat capacity normalized by the normal state value $C_{V}/ \gamma_N T$ as a function of $T/T_c$ (note that $T_c$ depends on $\G$). The columns correspond to different values of the pair breaking rate normalized by the value of the second gap, which is determined self-consistently $\Gamma/\Delta_{2}$. Note that the largest value of $\Gamma/\Delta_2$ is compared with the experimental data for the specific heat of 4Hb-TaS$_2$ from Ref.~\cite{ribak2020chiral}. }
	\label{fig:Main}
\end{figure*}

In the top row of Fig.~\ref{fig:Main} (a)-(d), we plot the total density of states determined self-consistently for varying $\G / \D_2$ and fixed $\nu_2 / \nu_1 = 0.15$, based on the experimental results on 4Hb-TaS$_2$ in Ref. \cite{ribak2020chiral}.  As $\G$ increases, states begin to fill the gap. When $\G \gg \D_2$ and Eq.~\eqref{eq:SmallXlim} holds, the density of states of band 2 near $\omega=0$ is almost constant, as in a metal. 
We note that this limit is robust, as it yields nearly identical results for any larger $\G$ value. Thus, the tunneling density of states is expected to resemble that of a fully gapped and relatively clean superconductor, but simply shifted by a constant. In particular, the coherence peak remains sharp, in contrast to the case of gapless superconductivity in single-band systems, where the coherence peak is suppressed in the gapless regime \cite{woolf1965effect}.
In the middle row of Fig.~\ref{fig:Main} (a)-(d), we plot the corresponding gap function Eq.~\eqref{eq:D} versus Matsubara frequency. The gap is compared with the Pad\'e approximation \cite{beach2000reliable} (still in Matsubara space) and the asymptotic form, Eq.~\eqref{eq:SmallXlim}. These plots demonstrate that the asymptotic form of the gap is a good approximation in the limit of strong disorder. 

Finally, we turn to the computation of the heat-capacity $C_{V}/T =-\dfrac{\partial^{2}{F}}  {{\partial T}^{2}}$, where $F =k_{B} T \int_{0}^{\infty} \ln \left[1-\exp(-\beta\omega) \right] \nu(\omega) d\omega $ is the free energy with Boltzmann's constant $k_{B}$ and $\beta = (k_{B}T)^{-1}$. Note that if $\Delta$ is only weakly dependent on the temperature, one can estimate the heat capacity by the following expression \cite{fernandes2011scaling}:
\begin{align}
{C_{V}\over T} \sim \int_{0}^{\infty}  \omega^{2}\left( -\dfrac{\partial f(\omega)}{\partial \omega}  \right) \nu(\omega) d\omega\,\;,\,\;T\to 0\,,
\end{align}
where $f(\omega) = (1+{\rm e}^{\omega/T})^{-1}$ is the Fermi-Dirac distribution function. At low temperatures ($T \lesssim T_{c}/2$), the results from this expression coincide with the numerical calculations.
In the bottom row of Fig.~\ref{fig:Main} (a)-(d), we plot the corresponding heat capacity normalized by the normal state contribution. Note that the numerical calculation of the gap and the analytic continuation become unstable close to the transition point. Therefore we plot the computed heat capacity only up to $T = T_c/2$. 
In column (d), we compare the results in the strong disorder regime with the experimental data from Ref.~\cite{ribak2020chiral}. Excellent agreement is found for the low-temperature behavior. This demonstrates that the experimental observation is consistent with our toy model's predictions.  

\begin{figure*}[h]
	\centering
	\includegraphics[width=0.9\textwidth]{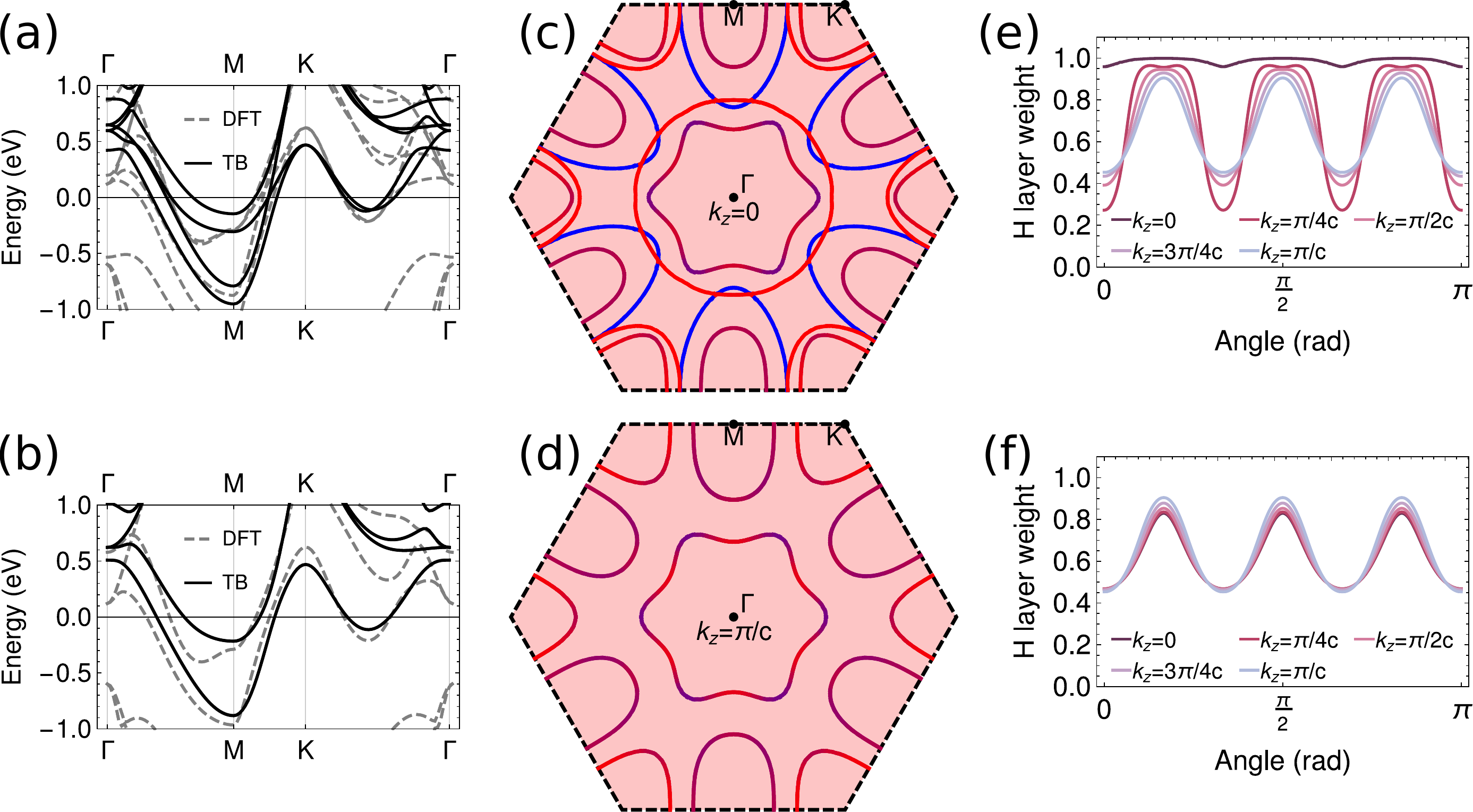}
	\caption{\label{fig:bands-compare} (a)-(b) Comparison of DFT (black, solid) and tight-binding (TB, gray, dashed) band structures for $k_z=0$ (a) and $k_z=\pm\pi/c$ (b). The two-fold degeneracy is clear for $k_z=\pm\pi/c$ and it is correctly captured by the tight binding model. (c)-(d) Fermi surfaces for $k_z=0$ and $k_z=\pm\pi/c$. They consist of pairs of pockets centered at the high-symmetry points of the Brillouin zone, $\G$, $M$, and $K$. These pairs become degenerate at $k_z = \pm \pi/c$ due to the screw symmetry of the crystal. The Fermi pockets are colored red (blue) according to their H-layer (T-layer) spectral weight. Therefore, purple means more strongly hybridized pockets. (e)-(f) Layer weights along outer (up) and inner (down) Gamma pockets. Outer pocket becomes strongly hybridized away from $k_z=0$.}
\end{figure*}

\section{Band structure of 4Hb-TaS$_2$ } \label{TBM}
After having established a simple toy model that demonstrates robust gapless superconductivity over a wide parameter range, we now discuss its relevance to 4Hb-TaS$_2$. Our first step is to construct a realistic tight-binding model from density-functional theory (DFT). We obtain a twelve-band tight binding model using DFT and maximally localized Wannier functions using the VASP code and the  projector-augmented-wave (PAW) approach \cite{VASP,VASP-PAW}. The exchange-correlation functional was approximated using the PBE potentials, which employ the generalized gradient approximation (GGA) \cite{Perdew1996Generalized}. As the unit cell is highly extended in the stacking direction, a $\Gamma$ centered $k$-points grid of 11x11x2 was used along with a plane wave cutoff of 470 eV. Structural parameters were taken from experiments \cite{disalvo73preparation}. 
The tight-binding model parameters were obtained by constructing maximally localized Wannier functions using the wannier90 package \cite{wannier90,Marzari2012Wannier}. It consists of 12 Ta orbitals, 3 per Ta atom per layer. The shape of these orbitals are different for atoms in the T-layer and in the H-layer (see Appendix \ref{sec:full-TBM}) with the T-layer having standard $t_{2g}$ orbitals and the H-layer having unconventional orbitals. The tight-binding parameters used here to reproduce the features of the Fermi surface and the low-energy band structure contain only onsite terms, nearest-neighbor hopping, and a single inter-layer hopping term. They can be found in Table \ref{tab:tb-params} in Appendix \ref{sec:full-TBM}.

The crystal structure of 4Hb-TaS$_2$ is invariant under a $6_3$ screw operation, which is equivalent to a 60$^\circ$ rotation around the $z$ axis, followed by a half-unit cell translation along the same axis. The structure is also invariant under a 120$^\circ$ degree rotation around $\hat{z}$, and the consecutive application of this rotation with the $6_3$ screw gives the $2_1$ screw operation, which we denote as $s$. $s$ is a 180$^\circ$ rotation followed by a half-unit cell translation.
The Bloch wave-function representation of the non-symmorphic screw-symmetry $s$ of the 4Hb compound squares to minus one, $s^2 = -1$, at $k_z=\pm\pi/c$. Due to this non-symmorphic symmetry, the band structure becomes doubly degenerate (four-fold including Kramers degeneracy) at all $k$-points on the $k_z=\pm\pi/c$. This can be seen by noting the relationships between the screw symmetry $s$, the mirror plane parallel to the H planes ($m_{001}$), and the inversion ($i$) with inversion centers at the T-layer Ta atoms. Because $s = m\cdot i$ and $m^2=i^2=1$, one would expect that $s^2=1$. But, as discussed above, $s^2=-1$ at $k_z=\pm\pi/c$. This apparent contradiction is resolved by enforcing a two-fold degeneracy throughout the $k_z=\pm\pi/c$ plane, with these degenerate wave-functions being related by $s$ \cite{dresselhaus2008applications, Hund1936}.

The resulting DFT band structures at $k_z=0$ and $k_z=\pm \pi/c$ are shown in Fig.~\ref{fig:bands-compare}(a)-(b), along with the tight-binding dispersions. The latter reproduces the correct number of distinct Fermi surfaces, shown in Fig.~\ref{fig:bands-compare}(c)-(d), which consist of pairs of split inner and outer pockets at $\Gamma$, $M$, and $K$ -- except at the $k_z=\pm \pi/c$ planes, where the inner and outer pockets become degenerate.

The Fermi pockets in Fig.~\ref{fig:bands-compare}(c)-(d) are colored according to their spectral weights, with red (blue) denoting spectral weight due to states from an H (T) layer. For $k_z=0$, while all inner pockets have a strong mixed-layer character (as indicated by their purple color), the outer pockets are made out of states from nearly a single layer only. In particular, the outer pockets at $\G$ and $K$ have a predominant H-layer character, whereas the outer pocket at $M$ has a dominant T-layer character. Thus, the H-T inter-layer hybridization is different for each of the Fermi pockets that form a pair of pockets centered at the same high-symmetry point of the Brillouin zone. As $k_z$ increases towards $k_z = \pm \pi/c$, the inter-layer hybridization becomes comparable for the pockets that form a pair. 

This behavior can be understood from the fact that the H-T hybridization between crossing bands depends on their mirror  eigenvalues. There is a mirror plane on each H layer, and Bloch states on the $k_z=0$ plane have a mirror eigenvalues of either $+1$ or $-1$. Symmetry dictates that the orbitals on the H layers induce bands with +1 mirror eigenvalues only, whereas the orbitals on T layers induce equal numbers of +1 and -1 eigenvalue bands. 
Hybridization between the + and - eigenvalue bands are forbidden on the $k_z=0$ plane, since these $k$ vectors don't break the mirror symmetry. As a result, the T pockets that have -1 mirror eigenvalue can cross the H pockets without hybridization on this plane. 

\section{Pair-breaking due to magnetic fluctuations on the T-layers}\label{Scattering-Rate}

We now connect the band structure calculations with our two-band toy model. Motivated by the fact that the 2H polymorph is a superconductor, whereas the 1T polymorph is a Mott insulator, we focus on the Fermi pockets of the 4Hb polymorph with dominant H-layer character, i.e. the pairs of pockets centered at $\G$ and $K$ in Fig.~ \ref{fig:bands-compare}(c). We further neglect large-momentum pairing interactions, and therefore consider each pair of pockets as an independent two-band model. In other words, the two bands in our toy model corresponds to the inner and outer Fermi pockets centered at the same high-symmetry point of the Brillouin zone ($\G$ or $K$).  

Our working hypothesis is that superconductivity would emerge in these two pockets intrinsically, driven by the same intra-pocket pairing interaction that makes the 2H polymorph a superconductor. The key difference in the 4Hb compound is that these bands hybridize with the T-layer states, as we discussed above. Now, the 1T polymorph is a Mott insulator, presumably with strongly fluctuating local moments that do not order magnetically. We assume that these local-moment fluctuations persist in the 4Hb compound, since it also undergoes a star-of-David CDW transition. These magnetic fluctuations on the T layer will then cause pair-breaking in the superconducting H-layer pockets via the H-T interlayer hybridization. Because the hybridization is significantly different for the two pockets, particularly near $k_z=0$, the pair-breaking effect is expected to be of very distinct magnitudes in each pocket.

A full description of the local moments in the T-layer and of their coupling to the H-layer itinerant states is well beyond the scope of this work. To capture the pair-breaking effect caused by these magnetic fluctuating moments, we model them as dilute magnetic impurities. We then compute the single-particle lifetime in the pairs of Fermi pockets centered at $\G$ and $K$ due to such impurities residing on the T-layers, in order to verify that the pair-breaking potential is significantly different in the inner and outer pockets centered at each momentum.

\begin{figure*}[tbh]
	\centering
	\includegraphics[width=0.85\textwidth]{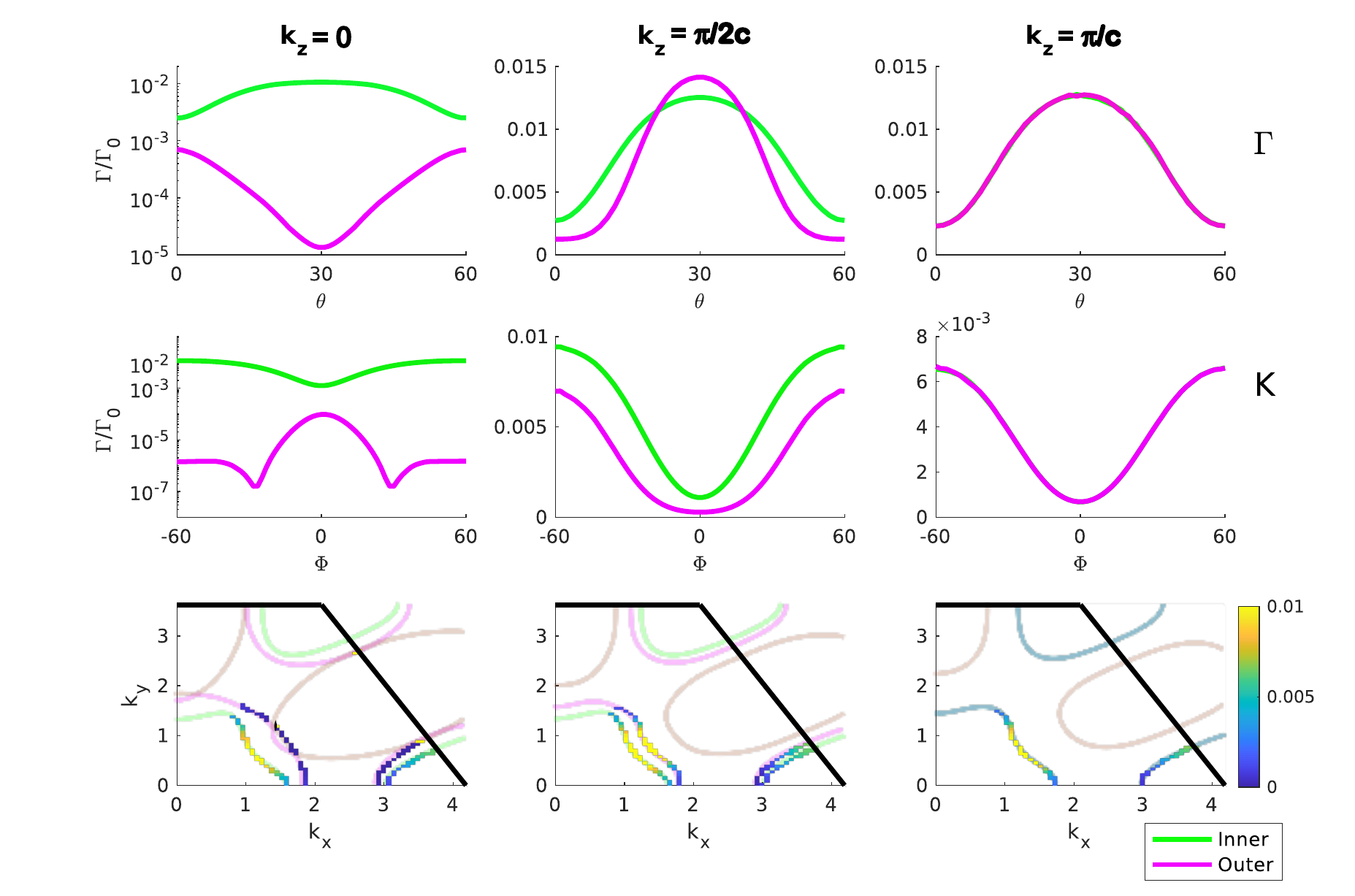}
	\caption{The inverse lifetime $\G$ due to scattering by impurities, Eq.~(\ref{GammaValue}), normalized by $\Gamma_{0} = n_{imp}V^2/\Omega W$, where $\Omega$ is the unit-cell volume and $W \sim 1$[eV] is comparable to the typical bandwidth. This quantity is a proxy for the pair-breaking scattering rate.
		The inverse lifetime is presented for different $k_{z}$ values for the inner (green) and outer (violet) pockets centered at the $\Gamma$ (upper panel) and  $K$ (middle panel) points. The angle is measured with respect to $k_{y}=0$, where $\theta$ starts at $(0, 0)$ counter clockwise, and $\Phi$ at $(4\pi/3,0)$ clockwise. At $k_{z} = 0$, the inverse lifetime on the outer pocket is orders of magnitude smaller than that in the inner one (note that in this case the plot is logarithmic in the y-axis). However, it grows rapidly and becomes equal to the latter as the zone boundary is approached. The lower panel shows the same inverse lifetime, overlaid on the corresponding Fermi pockets at the right corner of the Brillouin zone (marked by thick black lines). Here we can also observe a considerable difference between the $K$ pockets and the $\G$ pockets.}
	\label{fig:PairBreakingValue}
\end{figure*}

We start from the bare Green's function in the {\it band basis}
\begin{align}
G_{0}^j(i\omega,\bs{k}) = \dfrac{1}{-i\omega+(\varepsilon_{\bs{k},j}-\varepsilon_{F})},
\end{align}
where $\varepsilon_{\bs{k},j}$ are the eigenvalues of the 12$\times$12 tight-binding Hamiltonian (see Appendix \ref{sec:full-TBM}).

The disorder potential in the {\it orbital basis} is assumed to be point like and uncorrelated
\begin{align}\label{dis1}
S_{d} = \sum_{l=1}^{N_{imp}}\sum_{\omega, \bs{k},\bs{p}} V_{l} {\rm e}^{i(\bs{k}-\bs{p})\cdot \bs{r}_{l}}\psi^{\dagger}_{\omega,\bs{p}} M \psi_{\omega,\bs{k}},
\end{align}
where $N_{imp}$ is the number of impurities and $\bs{r}_{l}$ are their positions. The matrix $M$ encodes the orbital structure of the disorder potential. Following our convention of T-H-T-H stacking, we model the impurities in the T-layers by the matrix:
\begin{align}
M = \begin{pmatrix}
\mathbb{1} & 0 & 0 & 0 \\  0 & 0 & 0 & 0 \\ 0 &  0 & \mathbb{1} & 0 \\  0 & 0 & 0 & 0 
\end{pmatrix},
\end{align}
where $\mathbb{1}$ is a $3 \ \times \ 3$ identity matrix. For the purposes of computing the lifetime, which is only a proxy for the pair-breaking scattering rate, it is fine to consider such a non-magnetic potential, since the tight-binding model is SU(2) symmetric. 

We now rotate Eq.~(\ref{dis1}) to the band basis, such that the disorder potential assumes the form
\begin{align}
S_{d} = \sum_{l=1}^{N_{imp}}\sum_{\omega, \bs{k},\bs{p}} V_{l} {\rm e}^{i(\bs{k}-\bs{p})\cdot \bs{r}_{l}}c^{\dagger}_{\omega,\bs{p}} Q(\bs{p},\bs{k}) c_{\omega,\bs{k}},
\end{align}
where  $Q^{\alpha \beta}(\bs{p},\bs{q}) = \langle \bs{p} \alpha|M|\bs{k} \beta  \rangle$. 
Then the self-energy is given by 
\begin{align}
\Sigma(i\omega,\bs{k}) = \int_{BZ} \,d^{3}p \ Q(\bs{k},\bs{p}) {G}_{0}(i\omega, \bs{p}) Q(\bs{p},\bs{k}),
\end{align}
where the integration is over the entire Brillouin zone assuming a symmetric point-like disorder distribution~\cite{monkhorst1976special}. The inverse lifetime can then be calculated via:
\begin{align} \label{GammaValue}
\Gamma(\bs{k}) = \Im \left[\Sigma(i\omega \to 0^{+},\bs{k}) -\Sigma(i\omega \to 0^{-},\bs{k})\right].
\end{align}

The inverse lifetime $\G$ is plotted in Fig.~\ref{fig:PairBreakingValue} for three different $k_z$ planes: $k_{z} =0$, $k_z = \pm \pi/2c$, and $k_z = \pm \pi/c$. In particular, for each column, we show the inverse lifetime on the inner (green) and outer (violet) Fermi pockets centered at the $\Gamma$ and $K$ points (i.e. the Fermi pockets with predominant H character). 
We find that, at $k_z = 0$, the inverse lifetimes on the inner Fermi pockets are orders of magnitude larger than those on the outer pockets. However, as we move away from the $k_z = 0$ plane, the inverse lifetimes on the outer pockets grow rapidly and become nearly equal to those on the inner pockets. We also note that the scattering rate on the $K$ pockets is consistently smaller than that of $\G$ pockets. 

Based on these results, and using the inverse lifetime as a proxy for the pair-breaking scattering rate, we conclude that the scattering rate in 4Hb-TaS$_2$ depends strongly on the Fermi surface. This qualitatively justifies the toy model studied in Section \ref{Toy Model}. The difference  is greatest when scattering involving the $k_z = 0$ states provides the dominant pair-breaking channel.

\section{Discussion}\label{Discussion}
In this paper we proposed a simple scenario for the superconducting ground state of 4Hb-TaS$_{2}$. In particular, we showed that the puzzling $T$-linear specific heat that is observed experimentally at low temperatures is naturally explained when the pair-breaking scattering rate varies significantly between the different Fermi pockets of a two-band superconductor. In this situation, the well-known gapless superconducting state predicted by Abrikosov and Gor'kov~\cite{abrikosov1960contribution} is stabilized over a wide range of parameters. The presence of both a fully gapped superconducting pocket and a gapless superconducting pocket ensures the existence of sharp coherent peaks at the density of states and of a sharp specific heat jump at $T_c$.

To show the relevance of this toy model to 4Hb-TaS$_{2}$, we derived a tight-binding model for this compound from DFT. We find that the DFT band structure is well captured within a nearest-neighbour tight-binding approximation including only one (diagonal) interlayer hopping term. Importantly, we found that the special screw symmetry of this compound forces a degenerate doublet of Fermi pockets at the zone top and bottom, which are otherwise non-degenerate and form ``inner" and ``outer" Fermi pockets surrounding the $\Gamma$, $K$, and $M$ points. 

We then employed this tight-binding model to estimate the expected pair-breaking scattering rates on the different Fermi sheets with predominant H-layer character, caused by magnetic fluctuations residing on the strongly-correlated T-layers. We found the rates can be notably different when comparing the inner vs. outer Fermi surfaces. In particular,
near the $k_z = 0$ plane, the inner pockets experience a scattering rate that is more than two orders of magnitude larger than  that experienced by the outer pockets. 
Additionally, we also observe a considerable difference between the scattering rate on Fermi pockets surrounding $K$ and those surrounding $\G$ throughout the entire Brillouin zone. 
We note that pocket-dependent pair-breaking and multi-gap superconductivity  were recently  observed in the related compound 2H-NbSe$_2$~\cite{dvir2018spectroscopy}. Moreover, recent magneto-transport measurements performed on 4Hb-TaS$_2$ reveal a large variability of the mobility of carriers on the different Fermi pockets~\cite{gao2020origin}.

It is important to note that our tight-binding model neglects the effects of spin-orbit coupling and the CDW phases in both T and H layers. While we checked that the influence of the spin-orbit coupling on the band structure is weak compared to the impact of the inter-layer coupling, the CDW phases may have a more important effect. For example, it is possible that the H-T hybridization is significantly modified. Nevertheless, the conclusion that a Fermi-surface dependent scattering rate leads to robust gapless superconductivity is more general, and may even apply to other superconductors exhibiting anomalous specific heat behavior~\cite{ran2019nearly}. 
 
In this paper, we assumed that the magnetic disorder is static. However, it is interesting to consider the implications of dynamic fluctuations, in which the inner Fermi pockets may Kondo-screen them. We note that Kondo-screening between the H and T layers of TaSe$_2$ was recently reported~\cite{ruan2020imaging}. Such a screening may also occur inside the superconducting state, as the inner Fermi surfaces are gapless \cite{borkowski1992kondo}. 
In fact, it is possible that the Josephson coupling to the outer Fermi pockets competes (weakly) with the Kondo screening and causes a partial unscreening of the magnetic moments. We thus raise the question whether this may be a possible source for the enhanced muon relaxation observed below $T_c$, which would then be unrelated to chiral superconductivity.

We also note that such a Kondo screened state is not expected to leave signatures in transport because the gapped superconducting Fermi pockets shunt the gapless ones. To observe these effects, we propose the use of local probes with scanning capabilities such as scanning SQUID, STM and compressibility sensors. When the Kondo temperature is low, it may be possible to unscreen the magnetic moments in a magnetic field and observe an anomalous magnetic response even above the superconducting transition temperature.

\begin{acknowledgments}
	We are grateful to Patrick Lee and Vladyslav Kozii for helpful discussions. We owe special gratitude to Yoram Dagan and Amit Kanigel for sharing their data with us.
	JR and DD were funded by the Israeli Science Foundation under Grant No. 994/19. JR acknowledges the support of the Alon fellowship provided by the Israeli high council of education. 
	The work at the University of Minnesota (EDR, TB, and RMF) was supported by the National Science Foundation through the University of Minnesota MRSEC under Award Number DMR-2011401. We acknowledge the Minnesota Supercomputing Institute (MSI) at the University of Minnesota for providing resources that contributed to the research results reported within this paper. 
\end{acknowledgments}

\onecolumngrid
\appendix
\renewcommand\thefigure{\thesection.\arabic{figure}}

\section{Details of the Tight Binding Model}\label{sec:full-TBM}
\setcounter{figure}{0}

\begin{figure}
	\centering
	\includegraphics[width=0.7\textwidth]{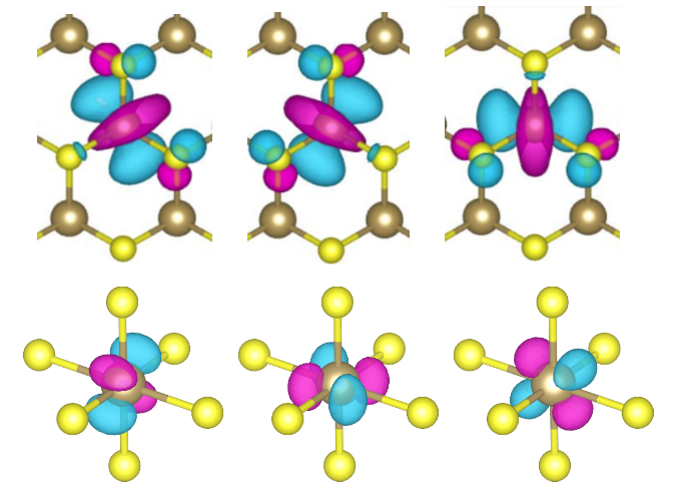}
	\caption{Maximally localized Wannier orbitals for H layer (upper panel) and T layer (lower panel)}
	\label{fig:orbitals}
\end{figure}

Our tight binding Hamiltonian has three orbitals per Ta atom, as presented in Fig.~\ref{fig:orbitals}, where we show the maximally localized Wannier orbitals used to construct it. 
The Wannier function calculations are initialized by using random projections; in other words, the shapes and symmetries of these orbitals are not imposed manually and are rather an outcome of the calculation. Even though the site-symmetry of the Ta sites on the T layer ($\bar{3}m$) is not cubic, the deviation from the cubic symmetry is not strong in the crystal field, and the resulting Wannier orbitals are very similar to the $d_{xy}$, $d_{xz}$, and $d_{yz}$ cubic harmonics with cartesian axes chosen parallel to the Ta-S bonds (Fig.~\ref{fig:orbitals}, bottom panels). 
For the H layer (upper panel), all three Wannier orbitals have shapes similar to the $d_{3z^2-r^2}$ orbital, but they are oriented in-plane and towards a neighboring pair of Ta atoms. The shape of these orbitals is a result of the local crystal field. The site-symmetry of the Ta site on the H layers is $\bar{6}m2$. The trigonal prismatic crystal field splits the otherwise 5-fold degenerate $d$ levels into a doublet and a singlet, where the energy splitting of the lower doublet and the singlet is $\sim100$~meV in TaS$_2$. For $z$ axis chosen normal to the H layer, the singlet, which transforms as the $A_1'$ irrep, has $d_{3z^2-r^2}$ character. Similarly, the lower doublet, which transforms at the $E'$ irrep, has the $(d_{xy}, d_{x^2-y^2})$ character. The following superpositions of these three cubic harmonic orbitals give 3 orbitals, similar to those in the upper panel of Fig.~\ref{fig:orbitals}, which transform to each other under 120$^\circ$ rotations: 
\begin{equation}
\begin{split}
d_1&=-\sqrt{\frac{1}{3}}d_{3z^2-r^2}+\sqrt{\frac{2}{3}}d_{x^2-y^2}\\
d_2&=-\sqrt{\frac{1}{3}}d_{3z^2-r^2}-\sqrt{\frac{1}{2}}d_{xy}-\sqrt{\frac{1}{6}}d_{x^2-y^2}\\
d_3&=-\sqrt{\frac{1}{3}}d_{3z^2-r^2}+\sqrt{\frac{1}{2}}d_{xy}-\sqrt{\frac{1}{6}}d_{x^2-y^2}
\end{split}
\label{equ:z2}
\end{equation}
These orbitals are orthonormal\footnote{A similar, but  not orthonormal, basis with 4 different $d_{3z^2-r^2}$-like orbitals was discussed in Ref.~\cite{VanWezel2012} for 2H-TaS$_2$. }. The Wannier functions don't have exactly the same form as in Eq.~\ref{equ:z2}, for they also include the hybridization between different orbitals and atoms. Also, note that the $d_{3z^2-r^2}$ orbital oriented along the $x$ direction, which we call $d_{3x^2-r^2}$, can be written as
\begin{equation}
d_{3x^2-y^2}=-\frac{1}{2}d_{3z^2-r^2}+\frac{\sqrt{3}}{2}d_{x^2-y^2} .
\end{equation}
Thus, the orbital character of $d_1$ is very similar to $d_{3x^2-r^3}$, except for the loss of continuous rotational symmetry around the $x$ axis, which does not exist in the crystal. 

In summary, we find that on every layer of TaS$2$, there are 3 different atom-centered Wannier orbitals that need to be included in the tight binding model. This leads to a $12\times12$ tight-binding Hamiltonian. We include onsite off-diagonal terms on the H layers, and select in-plane nearest neighbor hopping terms, in addition to one out-of-plane term. The layers are specified in stacking order T-H-T-H. The structure of the matrix is:

\begin{align}\label{TBMH}
\hat H = \left(
\begin{array}{cccccccccccc}
\xi_{1,1} & 0 & 0 & \xi_{1,4} & 0 & 0 & 0 & 0 & 0 & \xi_{1,10} & 0 & 0 
\\
0 & \xi_{2,2} & 0 & 0 & \xi_{2,5} & 0 & 0 & 0 & 0 & 0 & \xi_{2,11} & 0 
\\
0 & 0 & \xi_{3,3} & 0 & 0 & \xi_{3,6} & 0 & 0 & 0 & 0 & 0 & \xi_{3,12} 
\\
\xi_{1,4}^* & 0 & 0 & \xi_{4,4} & \xi_{4,5} & \xi_{4,6} & \xi_{4,7} & 0 & 0 & 0 & 0 & 0 
\\
0 & \xi_{2,5}^* & 0 & \xi_{4,5}^* & \xi_{5,5} & \xi_{5,6} & 0 & \xi_{5,8} & 0 & 0 & 0 & 0 
\\
0 & 0 & \xi_{3,6}^* & \xi_{4,6}^* & \xi_{5,6}^* & \xi_{6,6} & 0 & 0 & \xi_{6,9} & 0 & 0 & 0 
\\
0 & 0 & 0 & \xi_{4,7}^* & 0 & 0 & \xi_{7,7} & 0 & 0 & \xi_{7,10} & 0 & 0 
\\
0 & 0 & 0 & 0 & \xi_{5,8}^* & 0 & 0 & \xi_{8,8} & 0 & 0 & \xi_{8,11} & 0 
\\
0 & 0 & 0 & 0 & 0 & \xi_{6,9}^* & 0 & 0 & \xi_{9,9} & 0 & 0 & \xi_{9,12} 
\\
\xi_{1,10}^* & 0 & 0 & 0 & 0 & 0 & \xi_{7,10}^* & 0 & 0 & \xi_{10,10} & \xi_{10,11} & \xi_{10,12} 
\\
0 & \xi_{2,11}^* & 0 & 0 & 0 & 0 & 0 & \xi_{8,11}^* & 0 & \xi_{10,11}^* & \xi_{11,11} & \xi_{11,12} 
\\
0 & 0 & \xi_{3,12}^* & 0 & 0 & 0 & 0 & 0 & \xi_{9,12}^* & \xi_{10,12}^* & \xi_{11,12}^* & \xi_{12,12} 
\\
\end{array}
\right)
\end{align}

The matrix elements are given by:

\begin{align*}
\xi_{1,1} =& \epsilon_H-2 t_1 \cos( \vec{k} \cdot \vec{x}_3 )+2 t_2 [\cos( \vec{k} \cdot \vec{x}_1 )+\cos( \vec{k} \cdot \vec{x}_2 )]
\\
\xi_{1,4} =& -t_z \{i \sin[ \vec{k} \cdot (\vec{c}_1+\vec{x}_1) ]+ i\sin[ \vec{k} \cdot (\vec{c}_1-\vec{x}_2) ]+ \cos[ \vec{k} \cdot (\vec{c}_1+\vec{x}_1) ]+ \cos[ \vec{k} \cdot (\vec{c}_1-\vec{x}_2) ]\}
\\
\xi_{1,10} =& -t_z \{-i \sin[ \vec{k} \cdot (\vec{c}_1+\vec{x}_1) ]- i\sin[ \vec{k} \cdot (\vec{c}_1-\vec{x}_2) ]+ \cos[ \vec{k} \cdot (\vec{c}_1+\vec{x}_1) ]+ \cos[ \vec{k} \cdot (\vec{c}_1-\vec{x}_2) ]\}
\\
\xi_{2,2} =& \epsilon_H-2 t_1 \cos( \vec{k} \cdot \vec{x}_1 )+2 t_2 [\cos( \vec{k} \cdot \vec{x}_2 )+\cos( \vec{k} \cdot \vec{x}_3 )]
\\
\xi_{2,5} =& t_z \{i\sin[ \vec{k} \cdot (\vec{c}_1+\vec{x}_2) ]+ i\sin[ \vec{k} \cdot (\vec{c}_1+\vec{x}_3) ]+ \cos[ \vec{k} \cdot (\vec{c}_1+\vec{x}_2) ]+ \cos[ \vec{k} \cdot (\vec{c}_1+\vec{x}_3) ]\}
\\
\xi_{2,11} =& t_z \{-i\sin[ \vec{k} \cdot (\vec{c}_1+\vec{x}_2) ]- i\sin[ \vec{k} \cdot (\vec{c}_1+\vec{x}_3) ]+ \cos[ \vec{k} \cdot (\vec{c}_1+\vec{x}_2) ]+ \cos[ \vec{k} \cdot (\vec{c}_1+\vec{x}_3) ]\}
\\
\xi_{3,3} =& \epsilon_H-2 t_1 \cos( \vec{k} \cdot \vec{x}_2 ) + 2 t_2 [\cos( \vec{k} \cdot \vec{x}_1 )+\cos( \vec{k} \cdot \vec{x}_3 )]
\\
\xi_{3,6} =& -t_z \{i \sin[ \vec{k} \cdot (\vec{c}_1-\vec{x}_1) ]+ i\sin[ \vec{k} \cdot (\vec{c}_1-\vec{x}_3) ]+ \cos[ \vec{k} \cdot (\vec{c}_1-\vec{x}_1) ]+ \cos[ \vec{k} \cdot (\vec{c}_1-\vec{x}_3) ]\}
\\
\xi_{3,12} =& -t_z \{-i\sin[ \vec{k} \cdot (\vec{c}_1-\vec{x}_1) ]- i\sin[ \vec{k} \cdot (\vec{c}_1-\vec{x}_3) ]+ \cos[ \vec{k} \cdot (\vec{c}_1-\vec{x}_1) ]+ \cos[ \vec{k} \cdot (\vec{c}_1-\vec{x}_3) ]\}
\\
\xi_{4,4} =& \epsilon_T - 2 t_3 \cos( \vec{k} \cdot \vec{x}_3 )+ 2 t_4 [\cos( \vec{k} \cdot \vec{x}_1 )+\cos( \vec{k} \cdot \vec{x}_2 )]
\\
\xi_{4,5} =& t_o + t_6 [-i\sin( \vec{k} \cdot \vec{x}_1 ) +i \sin( \vec{k} \cdot \vec{x}_3 )+\cos( \vec{k} \cdot \vec{x}_1 )+\cos( \vec{k} \cdot \vec{x}_3 )] +
t_7 [i\sin( \vec{k} \cdot \vec{x}_2 )- \cos( \vec{k} \cdot \vec{x}_2 )]\\
& + t_8 [i\sin( \vec{k} \cdot \vec{x}_2 )+ \cos( \vec{k} \cdot \vec{x}_2 )]
-t_5 [i\sin( \vec{k} \cdot \vec{x}_1 )- i\sin( \vec{k} \cdot \vec{x}_3 )+\cos( \vec{k} \cdot \vec{x}_1 )+\cos( \vec{k} \cdot \vec{x}_3 )]
\\
\xi_{4,6} =& t_o 
+t_6 [i\sin( \vec{k} \cdot \vec{x}_2 )-i\sin( \vec{k} \cdot \vec{x}_3 )+ \cos( \vec{k} \cdot \vec{x}_2 )+\cos( \vec{k} \cdot \vec{x}_3 )]
-t_7 [i\sin( \vec{k} \cdot \vec{x}_1 )+\cos( \vec{k} \cdot \vec{x}_1 )]\\ 
&+ t_8[-i\sin( \vec{k} \cdot \vec{x}_1 )+ \cos( \vec{k} \cdot \vec{x}_1 )]
-t_5 [-i\sin( \vec{k} \cdot \vec{x}_2 )+i\sin( \vec{k} \cdot \vec{x}_3 )+\cos( \vec{k} \cdot \vec{x}_2 )+\cos( \vec{k} \cdot \vec{x}_3 )]
\\
\xi_{4,7} =& -t_z \{i \sin[ \vec{k} \cdot (\vec{c}_1-\vec{x}_1) ]+ i\sin[ \vec{k} \cdot (\vec{c}_1+\vec{x}_2) ]+ \cos[ \vec{k} \cdot (\vec{c}_1-\vec{x}_1) ]+ \cos[ \vec{k} \cdot (\vec{c}_1+\vec{x}_2) ]\}
\\
\xi_{5,5} =& \epsilon_T - 2 t_3 \cos( \vec{k} \cdot \vec{x}_1 ) +2 t_4 [\cos( \vec{k} \cdot \vec{x}_2 )+\cos( \vec{k} \cdot \vec{x}_3 )]
\\
\xi_{5,6} =& t_o +t_6 [-i \sin( \vec{k} \cdot \vec{x}_1 )-i\sin( \vec{k} \cdot \vec{x}_2 ) +\cos( \vec{k} \cdot \vec{x}_1 )+\cos( \vec{k} \cdot \vec{x}_2 )]
-t_7[i\sin( \vec{k} \cdot \vec{x}_3 )+\cos( \vec{k} \cdot \vec{x}_3 )]\\
&+t_8 [-i\sin( \vec{k} \cdot \vec{x}_3 )+\cos( \vec{k} \cdot \vec{x}_3 )]
-t_5 [i \sin( \vec{k} \cdot \vec{x}_1 )+\sin( \vec{k} \cdot \vec{x}_2 )+\cos( \vec{k} \cdot \vec{x}_1 )+\cos( \vec{k} \cdot \vec{x}_2 )]
\\
\xi_{5,8} =& t_z \{i \sin[ \vec{k} \cdot (\vec{c}_1-\vec{x}_2) ]+i\sin[ \vec{k} \cdot (\vec{c}_1-\vec{x}_3) ]+\cos[ \vec{k} \cdot (\vec{c}_1-\vec{x}_2) ]+\cos[ \vec{k} \cdot (\vec{c}_1-\vec{x}_3) ]\}
\\
\xi_{6,6} =& \epsilon_T - 2 t_3 \cos( \vec{k} \cdot \vec{x}_2 )+2 t_4 [\cos( \vec{k} \cdot \vec{x}_1 )+\cos( \vec{k} \cdot \vec{x}_3 )]
\\
\xi_{6,9} =& -t_z \{i \sin[ \vec{k} \cdot (\vec{c}_1+\vec{x}_1) ]+i\sin[ \vec{k} \cdot (\vec{c}_1+\vec{x}_3) ]+\cos[ \vec{k} \cdot (\vec{c}_1+\vec{x}_1) ]+\cos[ \vec{k} \cdot (\vec{c}_1+\vec{x}_3) ]\}
\\
\xi_{7,7} =& \epsilon_H-2 t_1 \cos( \vec{k} \cdot \vec{x}_3 )+2 t_2 [\cos( \vec{k} \cdot \vec{x}_1 )+\cos( \vec{k} \cdot \vec{x}_2 )]
\\
\xi_{7,10} =& -t_z \{i \sin[ \vec{k} \cdot (\vec{c}_1-\vec{x}_1) ]+i\sin[ \vec{k} \cdot (\vec{c}_1+\vec{x}_2) ]+\cos[ \vec{k} \cdot (\vec{c}_1-\vec{x}_1) ]+\cos[ \vec{k} \cdot (\vec{c}_1+\vec{x}_2) ]\}
\\
\xi_{8,8} =& \epsilon_H-2 t_1 \cos( \vec{k} \cdot \vec{x}_1 )+2 t_2 [\cos( \vec{k} \cdot \vec{x}_2 )+\cos( \vec{k} \cdot \vec{x}_3 )]
\\
\xi_{8,11} =& t_z \{i\sin[ \vec{k} \cdot (\vec{c}_1-\vec{x}_2) ]+i\sin[ \vec{k} \cdot (\vec{c}_1-\vec{x}_3) ])+\cos[ \vec{k} \cdot (\vec{c}_1-\vec{x}_2) ]+\cos[ \vec{k} \cdot (\vec{c}_1-\vec{x}_3) ]\}
\\
\xi_{9,9} =& \epsilon_H-2 t_1 \cos( \vec{k} \cdot \vec{x}_2 )+2 t_2 [\cos( \vec{k} \cdot \vec{x}_1 )+\cos( \vec{k} \cdot \vec{x}_3 )]
\\
\xi_{9,12} =& -t_z \{i\sin[ \vec{k} \cdot (\vec{c}_1+\vec{x}_1) ]+ i\sin[ \vec{k} \cdot (\vec{c}_1+\vec{x}_3) ]+ \cos[ \vec{k} \cdot (\vec{c}_1+\vec{x}_1) ]+ \cos[ \vec{k} \cdot (\vec{c}_1+\vec{x}_3) \}
\\
\xi_{10,10} =& \epsilon_T - 2 t_3 \cos( \vec{k} \cdot \vec{x}_3 )+2 t_4 [\cos( \vec{k} \cdot \vec{x}_1 )+\cos( \vec{k} \cdot \vec{x}_2 )]
\\
\xi_{10,11} =& t_o +t_6 [i \sin( \vec{k} \cdot \vec{x}_1 )-i\sin( \vec{k} \cdot \vec{x}_3 )+\cos( \vec{k} \cdot \vec{x}_1 )+\cos( \vec{k} \cdot \vec{x}_3 )]
-t_7 [i\sin( \vec{k} \cdot \vec{x}_2 )+\cos( \vec{k} \cdot \vec{x}_2 )]\\
&+t_8 [ -i\sin( \vec{k} \cdot \vec{x}_2 )+\cos( \vec{k} \cdot \vec{x}_2 )]
-t_5 [-i\sin( \vec{k} \cdot \vec{x}_1 )+ i\sin( \vec{k} \cdot \vec{x}_3 )+\cos( \vec{k} \cdot \vec{x}_1 )+\cos( \vec{k} \cdot \vec{x}_3 )]
\\
\xi_{10,12} =& t_o +t_6 [-i \sin( \vec{k} \cdot \vec{x}_2 ) + i\sin( \vec{k} \cdot \vec{x}_3 )+\cos( \vec{k} \cdot \vec{x}_2 )+\cos( \vec{k} \cdot \vec{x}_3 )]
+ t_7 [i\sin( \vec{k} \cdot \vec{x}_1 )- \cos( \vec{k} \cdot \vec{x}_1 )]\\
&+ t_8 [i\sin( \vec{k} \cdot \vec{x}_1 )+ \cos( \vec{k} \cdot \vec{x}_1 )] 
-t_5 [i\sin( \vec{k} \cdot \vec{x}_2 )-i\sin( \vec{k} \cdot \vec{x}_3 )+\cos( \vec{k} \cdot \vec{x}_2 )+\cos( \vec{k} \cdot \vec{x}_3 )]
\\
\xi_{11,11} =& \epsilon_T - 2 t_3 \cos( \vec{k} \cdot \vec{x}_1 )+2 t_4 [\cos( \vec{k} \cdot \vec{x}_2 )+\cos( \vec{k} \cdot \vec{x}_3 )]
\\
\xi_{11,12} =& t_o + t_6 [i \sin( \vec{k} \cdot \vec{x}_1 )+i\sin( \vec{k} \cdot \vec{x}_2 )+\cos( \vec{k} \cdot \vec{x}_1 )+\cos( \vec{k} \cdot \vec{x}_2 )]
+ t_7[i\sin( \vec{k} \cdot \vec{x}_3 )- \cos( \vec{k} \cdot \vec{x}_3 )]\\
&+ t_8 [i\sin( \vec{k} \cdot \vec{x}_3 )+ \cos( \vec{k} \cdot \vec{x}_3 )]
-t_5 [-i \sin( \vec{k} \cdot \vec{x}_1 )-i\sin( \vec{k} \cdot \vec{x}_2 )+\cos( \vec{k} \cdot \vec{x}_1 )+\cos( \vec{k} \cdot \vec{x}_2 )]
\\
\xi_{12,12} =& \epsilon_T - 2 t_3 \cos( \vec{k} \cdot \vec{x}_2 )+2 t_4 [\cos( \vec{k} \cdot \vec{x}_1 )+\cos( \vec{k} \cdot \vec{x}_3 )]
\end{align*}

Here, $\vec{x}_{1,2,3}$ are the three nearest neighbor vectors and $\vec{c}_1$ is the out-of-plane interlayer vector given by:

\begin{align}
\vec{x}_1 = a\hat x, \quad
\vec{x}_2 = a\left( -\frac{1}{2}\hat x + \frac{\sqrt{3}}{2} \hat y \right), \quad
\vec{x}_3 = a   \left( \frac{1}{2}\hat x + \frac{\sqrt{3}}{2} \hat y \right), \quad
\vec{c}_1 = \frac{c}{4} \hat z.
\end{align}

The onsite terms and hopping parameters are given in Table.~\ref{tab:tb-params}.

\begin{table}[htp]
	\begin{tabular}{ r r | c }
		\hline
		Onsite terms: & \\
		$\epsilon_H$ & 7.363 \\
		$\epsilon_T$ & 6.577 \\
		$t_o$ & -0.20 \\
		\hline
		Nearest neighbors terms: & & \\
		$t_1$ & 0.742 \\
		$t_2$ & 0.180 \\
		$t_3$ & 0.655 \\
		$t_4$ & 0.295 \\
		$t_5$ & 0.453 \\
		$t_6$ & 0.275 \\
		$t_7$ & 0.146 \\
		$t_8$ & 0.113 \\
		\hline
		Interlayer term: & \\
		$t_z$ & 0.081 \\
		\hline
	\end{tabular}
	\caption{\label{tab:tb-params}
		Tight-binding parameters grouped by hopping matrix elements.}
\end{table}

\section{Heat capacity data and nodal superconductivity}\label{app:nodal}
\setcounter{figure}{0}    

The working hypothesis of this paper is that the specific heat data is inconsistent with a nodal order parameter, even one that is lurking behind the $T$-linear contribution. In this appendix we show that the specific heat is indeed inconsistent with such a nodal state, or at least that the nodal state is below the noise level of the experiment.  To show this, we consider the temperature dependence of the derivative ${d\over dT}\left({C_V / T}\right)$ near $T = 0$. In the case of line nodes we expect 
\[\text{line nodes : }{d\over dT}\left({C_V \over T}\right) \sim \rm{const} \;,\; T \to 0\,,\]
while for point nodes
\[\text{point nodes : }{d\over dT}\left({C_V \over T}\right) \sim T \;,\; T \to 0\,.\]

\begin{figure}[h!]
	\centering
	\includegraphics[width=0.325\textwidth]{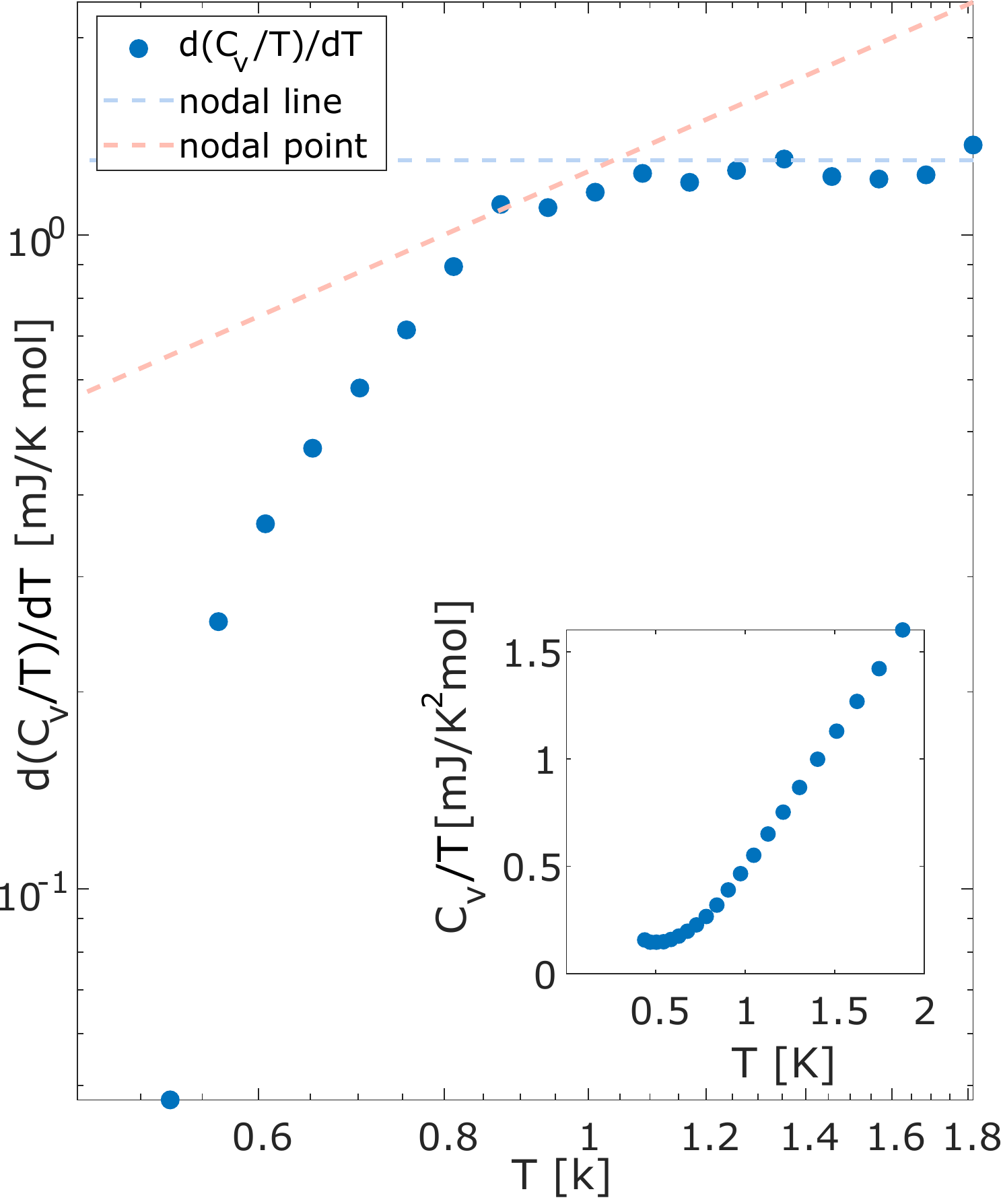}
	\caption{\label{fig:scrutny}   The derivative ${d\over dT}\left({C_V \over T}\right)$ as a function of $T$ on a log-log plot. The inset shows the raw data. 
	}
\end{figure}

In Fig.~\ref{fig:scrutny} we plot the numerical derivative
${d\over dT}\left({C_V \over T}\right)$
as a function of $T$ on a log-log plot for the experimentally measured specific heat data. It is clear that in the temperature range below $T = 1$K, the derivative is suppressed much faster than linear and is thus inconsistent with any kind of node. We thus conclude that $C_V$ most likely decays exponentially to a constant, or at least, in the temperature range down to 0.3K.

\twocolumngrid

\end{document}